\documentclass[10pt, prb, superscriptaddress, twocolumn]{revtex4-1}
\usepackage{graphicx}
\graphicspath{{./figs/}{./figures_tmp/}}
\usepackage{bm}
\usepackage{amssymb}
\usepackage{amsmath} 

\usepackage{xcolor}
\usepackage[bookmarks,urlbordercolor=white,]{hyperref}

\begin{document}
\title{Nano-scale magnetic skyrmions and target states in confined geometries}


\author{David Cort\'{e}s-Ortu\~{n}o}
\email{d.cortes@soton.ac.uk}
\affiliation{Faculty of Engineering and Physical Sciences, University of Southampton, Southampton SO17 1BJ, United Kingdom}
\author{Niklas Romming}
\affiliation{Institute of Applied Physics, University of Hamburg, Jungiusstrasse 11, D-20355 Hamburg, Germany}
\author{Marijan Beg}
\affiliation{European XFEL GmbH, Holzkoppel 4, 22869 Schenefeld, Germany}
\author{Kirsten von Bergmann}
\affiliation{Institute of Applied Physics, University of Hamburg, Jungiusstrasse 11, D-20355 Hamburg, Germany}
\author{Andr\'e Kubetzka}
\affiliation{Institute of Applied Physics, University of Hamburg, Jungiusstrasse 11, D-20355 Hamburg, Germany}
\author{Ondrej Hovorka}
\affiliation{Faculty of Engineering and Physical Sciences, University of Southampton, Southampton SO17 1BJ, United Kingdom}
\author{Hans Fangohr}
\affiliation{Faculty of Engineering and Physical Sciences, University of Southampton, Southampton SO17 1BJ, United Kingdom}
\affiliation{European XFEL GmbH, Holzkoppel 4, 22869 Schenefeld, Germany}
\author{Roland Wiesendanger}
\affiliation{Institute of Applied Physics, University of Hamburg, Jungiusstrasse 11, D-20355 Hamburg, Germany}


\begin{abstract}

	Research on magnetic systems with broken inversion symmetry has been
	stimulated by the experimental proof of particle-like configurations known as
	skyrmions, whose non-trivial topological properties make them ideal candidates
	for spintronic technology. In this class of materials, Dzyaloshinskii-Moriya
	interactions (DMI) are present, which favor the stabilization of chiral
	configurations. Recent advances in material engineering have shown that in
	confined geometries it is possible to stabilize skyrmionic configurations at
	zero field. Moreover, it has been shown that in systems based on Pd/Fe bilayers
	on top of Ir(111) surfaces skyrmions can be as small as a few nanometres in
	diameter. In this work we present scanning tunneling microscopy measurements of
	small Pd/Fe and Pd$_2$/Fe islands on Ir(111) that exhibit a variety of
	different spin textures, which can be reproduced using discrete spin
	simulations. These configurations include skyrmions and skyrmion-like states
	with extra spin rotations such as the target state, which have been of interest
	due to their promising dynamic properties. Furthermore, using simulations we
	analyze the stability of these skyrmionic textures as a function of island
	size, applied field and boundary conditions of the system. An understanding of
	the parameters and conditions affecting the stability of these magnetic
	structures in confined geometries is crucial for the development of
	energetically efficient and optimally sized skyrmion-based devices.

\end{abstract}


\maketitle



\section{Introduction}
\label{sec:intro}

Recent advances on the study and fabrication of ferromagnetic systems with
broken inversion symmetry have been highly motivated by the experimental
observation of topologically non-trivial and spatially localized particle-like
magnetic configurations known as skyrmions. The theoretical predictions of
skyrmions and skyrmion lattices in ferromagnetic materials with Dzyaloshinskii-Moriya interactions
(DMI)\cite{Bogdanov1989a,Bogdanov1989,Rossler2006} were first confirmed by M\"uhlbauer {\it et al.}~\cite{Muehlbauer2009} and later
in numerous experimental studies where non-collinear structures were
observed by imaging the magnetic texture of a sample.  In
particular, scanning tunneling microscopy with a spin-polarized
tip~\cite{Heinze2011,Romming2013,Romming2015,Wiesendanger2016} (SP-STM) has
been effective for sampling, with atomic-scale resolution, the presence and
structure of individual skyrmions in interfacial systems. Studies based on the SP-STM
technique applied to Pd/Fe films on an Ir(111) surface, where
skyrmions need to be stabilized using an external magnetic field, have shown
that individual skyrmions can be created and annihilated~\cite{Romming2013} and
how their size and shape depends on the field strength.~\cite{Romming2015}  Besides
skyrmions, other non-collinear structures have been experimentally observed as
equilibrium states. For example, spin spirals arise as the ground state in Pd/Fe
bilayers at zero field~\cite{Romming2013} and experiments show phases where
skyrmions coexist with spin spiral configurations. Furthermore, the theory predicts
different symmetrical configurations similar to skyrmions with additional
rotations~\cite{Bogdanov1994,Bogdanov1999}, which are known in the literature
as $k\pi$-skyrmions, with $k\in\mathbb{N}$ according to the number of
rotations.

A $2\pi$-skyrmion corresponds to a particle-like configuration with two
radially symmetric spin windings and is also known in the literature as the target
state.

Target states have been predicted by theory~\cite{Bogdanov1994,Bogdanov1999}
and observed in
micromagnetic~\cite{Rohart2013,Leonov2014a,Beg2015,Liu2015a,Carey2016,Pepper2018,Kolesnikov2018}
and atomistic~\cite{Hagemeister2018} simulations. It has been claimed that the
dynamics of target states caused by spin waves~\cite{Shen2018,Li2018},
field gradients~\cite{Komineas2015a} or spin-polarized
currents~\cite{Komineas2015,Liu2015a,Zhang2016,Kolesnikov2018} present some
differences with that of skyrmions, such as keeping a steady motion after
switching off the current. Other effects make evident their advantage over
skyrmions for their manipulation in two dimensional systems, such as racetrack
geometries. These include not being affected by the skyrmion Hall
effect,~\cite{Nagaosa2013} because their net topological charge is zero,
reaching larger velocities under currents applied perpendicular to the hosting
material,~\cite{Zhang2016,Kolesnikov2018} and not exhibiting distortion under
certain conditions during their motion.~\cite{Zhang2016,Komineas2015a} Although
recent experiments have reported the observation of target states in the
ferrimagnetic~\cite{Finazzi2013} $\text{Tb}_{22}\text{Fe}_{69}\text{Co}_{9}$,
the $\text{Ni}_{80}\text{Fe}_{20}$ ferromagnet next to a topological
insulator,~\cite{Zhang2018} and the chiral magnet~\cite{Zheng2017} FeGe, thus
far no evidence has been provided for thin ferromagnetic layers next to a heavy
metal, where interfacial DMI is present.

Experimental studies of chiral materials have recently focused on the analysis
of skyrmionic textures in confined geometries.~\cite{Zheng2017,Jin2017,Zhang2016,Ho2019}
Theoretical investigations have shown that in confined geometries skyrmions and
target states can be stabilized at zero field,~\cite{Beg2015,Pepper2018} which is of significant
importance for the potential design of skyrmion-based spintronic
devices.~\cite{Wiesendanger2016,Fert2017}

A notable feature of skyrmions in interfacial Pd/Fe samples is their
small size of only a few nanometers in radius, which can be an important
step forward towards the miniaturization of magnetic technology. This is in contrast
with skyrmions in bulk materials such as FeGe, where the helical length is
about $70\,\text{nm}$. The same holds for target states, which were observed in $160\,\text{nm}$
diameter cylinders.~\cite{Zheng2017} Although the stabilization of skyrmionic
textures in extended Pd/Fe bilayer samples requires the application of a magnetic
field, confined geometries offer an alternative for observing these
structures at zero or weak field strengths. It is important to notice that at
this low field regime, spiral structures are
energetically favored over skyrmions since the DMI and exchange energies
dominate over the Zeeman energy. Hence, it is important to gain an
understanding of ranges of parameters where spin spirals, skyrmions and
target states are stable for their potential observation in small confined
interfacial systems.

In this work we show through SP-STM measurements and numerical simulations,
that a variety of chiral configurations can be stabilized in small
Pd/Fe/Ir(111) islands of about $20\,\text{nm}$ in diameter. We reproduce the
experimental images using simulations based on a discrete spin model. We
validate the simulations of the islands by comparing the tilting of spins at
the boundary with the experimental findings. Simulations show that a variety of the experimentally observed
chiral configurations are accessible through a magnetic field sweep and we characterize them through their topological number.
Furthermore, we show that the confinement within the islands allows the
stabilization of skyrmion-like configurations, such as a target state or a
$3\pi$-skyrmion, from zero field up to a wide range of magnetic fields below $1\,$T.

In addition to the analysis of Pd monolayer islands on Fe/Ir(111), we show STM
images of Pd double layer islands on Fe/Ir(111) with the presence of a
configuration resembling a target state. In this system the environment of the
islands in applied field is ferromagnetic owing to the surrounding
field-polarized Pd/Fe layer. To understand the stability of skyrmions and
target states under different boundary conditions we perform a systematic study
of these configurations in perfectly shaped hexagonal islands under different
conditions. In this context, we compute the skyrmion and target state size and
energy as a function of island size, applied field, and boundary condition.
Moreover, we vary these parameters to calculate the ground states of the
system, which reveals conditions for the stability of different $k\pi$-skyrmion
states.

As an additional proof for the stability of skyrmions and target states,
we calculate the energy barrier separating a skyrmion from the
uniform ordering and the barrier between a target state and a skyrmion. We
compute the barriers using the Geodesic Nudged Elastic Band Method
(GNEBM)~\cite{Bessarab2015} which has been previously used to compute
transitions with the least energy cost between equilibrium configurations in
finite chiral systems.~\cite{Cortes2017,Stosic2017,Bessarab2018,Hagemeister2018}
Our results show that the stability of target states benefits from
ferromagnetic boundaries, in agreement with our experimental findings.

We start this paper by introducing in Sec.~\ref{sec:exp-field-sweep} results
obtained with SP-STM measurements on the Pd monolayer islands on Fe/Ir(111) 
together with the theoretical basis for the simulation of these systems.
Consequently, in Sec.~\ref{sec:simulations-islands} we show discrete spin
simulations of the experimentally fabricated quasi-hexagonal hcp Pd islands on
fcc Fe. Using different initial states we reproduce the field sweep experiment
in these islands in Sec.~\ref{sec:sim-field-sweep}. In
Sec.~\ref{sec:experiment-PdPdFeIr} we discuss additional experiments performed
on Pd islands on an extended Pd/Fe film on Ir(111) and simulations of these
samples with ferromagnetic boundary conditions.  To obtain an understanding of
the experimental observation of magnetic orderings resembling a target state in
the Pd$_2$/Fe islands, in Sec.~\ref{sec:sim-hexagons} we model perfectly
hexagonal islands and study the energy and size of skyrmions and target states
as a function of hexagon size, applied field and boundary conditions. In
addition, we characterize the lowest energy states with a full phase diagram of the
hexagon system in Sec.~\ref{sec:phase-diagram}. Finally, in
Sec.~\ref{sec:stability-sk-hexagons} we study the stability of skyrmions and
target states by means of energy barrier and energy path calculations with the
GNEBM.


\begin{figure*}[p!]
    \includegraphics[width=\textwidth]{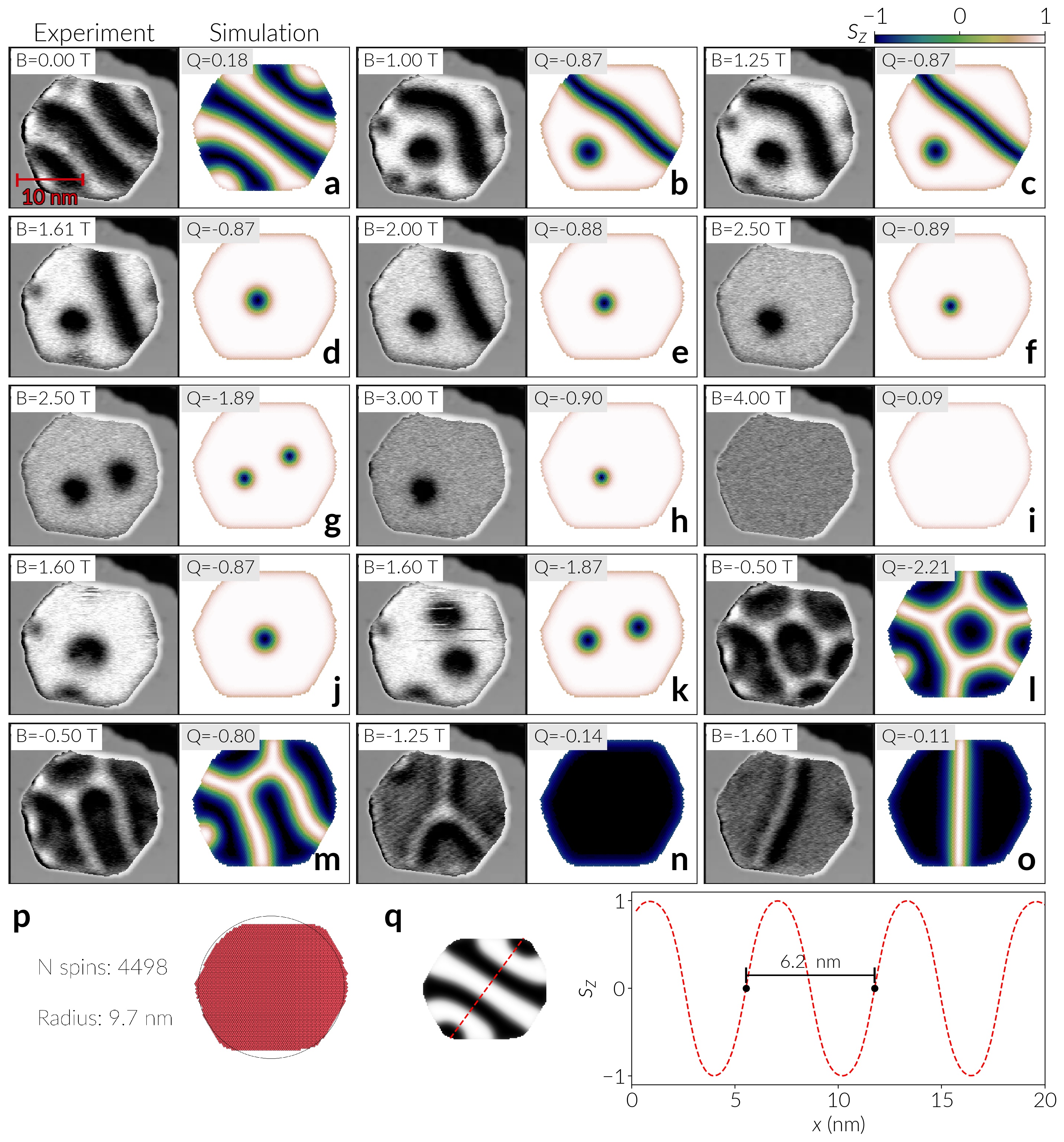}

    \caption{Magnetic states of a roughly hexagonal hcp-Pd island on
        fcc-Fe/Ir(111) during a magnetic field sweep. (a)-(o) Left images:
        spin-resolved differential conductance maps of a hexagonal Pd/Fe island
        measured with a Cr-bulk tip, bias voltage $U=+610\,\text{mV}$, the
        applied out-of-plane magnetic field $B$ is indicated; the image is
        composed of the differential conductance of the island together with
        the surrounding topography data. The magnetic tip is predominantly
        sensitive to the out-of-plane magnetization component.  Right images:
        corresponding discrete spin simulations, which are obtained by
        relaxation of the experimentally found magnetic state using the
        experimentally determined material parameters, see text; the calculated
        topological charge $Q$ is indicated. (p) Island shape used in the
        simulation.  (q) Out-of-plane component of the magnetization of the
        zero field spin spiral from the simulation in~(a), and a line profile
 		taken as indicated by the dashed line.  }

\label{fig:Romming-field-sweep}
\end{figure*}

\section{Magnetic states in Pd/Fe/Ir(111) islands}
\label{sec:exp-field-sweep}

\subsection{Experimentally observed magnetic states}
\label{sec:exp-obs-mag-states}

Spin-polarized scanning tunneling microscopy (SP-STM) is a valuable tool to
study nano-scale magnetism in model-type systems.~\cite{Wiesendanger2009}
Figure~\ref{fig:Romming-field-sweep} shows spin-resolved maps of the
differential conductance (left columns) of a monolayer thick hcp-stacked Pd
island on an extended fcc-Fe monolayer on Ir(111) at different applied
out-of-plane magnetic fields as indicated (cf.\ Supplemental Material, Sec.~S1,
for sample preparation~\cite{suppmat}). The presented data is obtained at a bias voltage of
$U=+610\,\text{mV}$ and the observed signal changes within the roughly
hexagonal island are due to a combination of tunnel magnetoresistance and
non-collinear magnetoresistance~\cite{Wiesendanger2009,Hanneken2015} (cf.\
Supplemental Material Sec.~S2 for more details~\cite{suppmat}). In this case the tip is
predominantly sensitive to the out-of-plane magnetization component of the
sample. The Pd-covered Fe (Pd/Fe bilayer) behaves as one magnetic entity: the
magnetic state of the Fe is changed by the hybridization with the Pd layer and
the Pd becomes spin-polarized. The Pd/Fe bilayer exhibits a spin spiral ground state
at zero magnetic field, as evident from the stripes observed in the
differential conductance map in (a).~\cite{Romming2013} Furthermore, a Pd/Fe
island is surrounded by the Fe monolayer, where a highly non-collinear
nano-skyrmion lattice~\cite{Heinze2011} is present. This nano-skyrmion lattice
does not change its state upon the application of an external magnetic field
and has a magnetic period of about 1~nm. As this is small compared to the
characteristic spin spiral wavelength of $6-7$~nm in Pd/Fe/Ir(111), we have
neglected a possible weak coupling between these layers in the theoretical
model, that we describe later, treating the island edges as free boundaries.

The investigated hexagonal island has a circumradius of about $9.7\,\text{nm}$
(see (p)) and the zero magnetic field spin spiral period is about 6~nm. This
spiral adjusts to the shape of the island by bending the stripes to be more
perpendicular to the island edges, see (a). When a magnetic field is applied,
the sample magnetization changes and at intermediate magnetic fields we observe
a skyrmion next to a $360^\circ$ wall, whereas at larger magnetic fields only
skyrmions are observed, see Fig.~\ref{fig:Romming-field-sweep}(b)-(h). Because
of the injection of higher energy electrons at the used bias voltage the tunnel
current can induce changes in the sample magnetization, compare (f) and (g),
which are measured at the same applied magnetic field but one skyrmion has
appeared between the two measurements. At $4\,\text{T}$ the sample reaches a
uniform magnetization, see Fig.~\ref{fig:Romming-field-sweep}(i). When the
applied magnetic field is reduced again similar magnetic states reappear, i.e.\
at $B=1.6\,\text{T}$ one or two skyrmions are found in the island, see (j)
and~(k).

However, when the applied magnetic field is then increased in the opposite
direction the magnetic configuration of the island is dominated by a network of
$360^\circ$ domain walls and their junctions, see
Fig.~\ref{fig:Romming-field-sweep}(l)-(n). These states are unusual magnetic
configurations and we think they arise when a magnetic field is applied to a
zero-field state that is not the virgin state but a metastable remanent state.
We would like to point out the central region of (l), which resembles a
magnetic skyrmion with the center (dark) parallel to the applied magnetic field
and the surrounding (bright ring) magnetized opposite to the applied magnetic
field; this means that this skyrmion has an equivalent spin texture as the one
observed in (f), but it exists here in the opposite magnetic field. When the
absolute value of the applied magnetic field is increased, more and more
junctions of $360^\circ$ walls disappear until $B=-1.6\,\text{T}$ when only one
$360^\circ$ domain wall remains; this is a state similar to the one found when
the magnetic field is applied to the virgin state, see
Fig.~\ref{fig:Romming-field-sweep}(d), which shows a similar $360^\circ$ domain
wall next to a skyrmion.

To obtain a better understanding of the magnetic field dependent stability and
topological properties of the observed magnetic states in this particular Pd/Fe
island, we performed discrete spin simulations of each magnetic state at the
given applied magnetic field using the Fidimag~\cite{Bisotti2018} code. We
translate the quasi hexagonal experimental Pd/Fe/Ir(111) island into a spin
lattice, as illustrated in Fig.~\ref{fig:Romming-field-sweep}(p). We treat the
Pd/Fe bilayer as a single entity and consequently simulate a monolayer of spins
in a two-dimensional triangular lattice. The islands are simulated with open
boundaries because of the Fe/Ir environment of the island explained earlier in
this Section. By setting the island in the $xy$-plane we use the following
Heisenberg-like Hamiltonian,

\begin{align}
    \mathcal{H}  = &
    - J \sum_{\langle i,j \rangle}^{P} \mathbf{s}_i \cdot \mathbf{s}_j
    + D\sum_{\langle i,j \rangle}^{P} \left( \mathbf{r}_{ij}\times\hat{\mathbf{z}} \right)
      \cdot \left[ \mathbf{s}_i \times \mathbf{s}_j \right] \nonumber \\
      & 
      - \mathcal{K} \sum_{i}^{P}
      \left( \mathbf{s}_i \cdot \hat{\mathbf{z}} \right)^{2}
      - \sum_{i}^{P} B_{z} \mu^{(z)}_i \,.
\label{eq:pdfeir-hamiltonian}
\end{align}

In Eq.~\ref{eq:pdfeir-hamiltonian} $\bm{\mu}_{i}=\mu_{i}\mathbf{s}_{i}$ is the
magnetic moment of the spin at the lattice site $i$, $\mathbf{s}_{i}$ is the
corresponding spin orientation with $|\mathbf{s}_{i}|=1$, $P$ is the number of
spins, which depends on the lattice constant and the size of the island,
$\mathcal{K}$ is an effective anisotropy taking into account the uniaxial
anisotropy and the approximation of the dipolar interactions, $D$ is the DMI
constant that describes the interfacial DMI for this material and the last term
is the Zeeman interaction for the field applied perpendicular to the sample
plane.

For the hcp stacking of Pd on fcc-Fe, as in this island, the continuum magnetic
parameters $A$, $D_{\text{c}}$, and $K_{\text{eff}}$ (exchange stiffness, DMI,
and effective uniaxial anisotropy) have been obtained from previous
experiments~\cite{Romming2015}, hence we convert these values into the
equivalent discrete magnitudes (see Sec.~S5 in the Supplemental Material for
details~\cite{suppmat}). These parameters define a characteristic helical
length~\cite{Leonov2016} at zero field and zero anisotropy of $L_{D}=4\pi
A/D_{\text{c}}=2\pi a J/D\approx 6.4\,\text{nm}$, where $a$ is the in-plane
atomic spacing of PdFe.
As in the previous work~\cite{Romming2015} we approximate dipolar
interactions into the anisotropy since, in the ultrathin film limit, their energy
contribution become significantly small~\cite{Leonov2016,Lobanov2016,Wang2018}
and it is possible to approximate them as a uniaxial
anisotropy,~\cite{Malottki2017} in particular for axisymmetric solutions (which
can be proved analytically~\cite{Bogdanov1994}). Note that if the stray field
is taken explicitly as in Ref.~\onlinecite{Kiselev2011} the anisotropy constant
is likely to have a smaller magnitude.

\begin{figure}[t!]
    \includegraphics[width=\columnwidth]{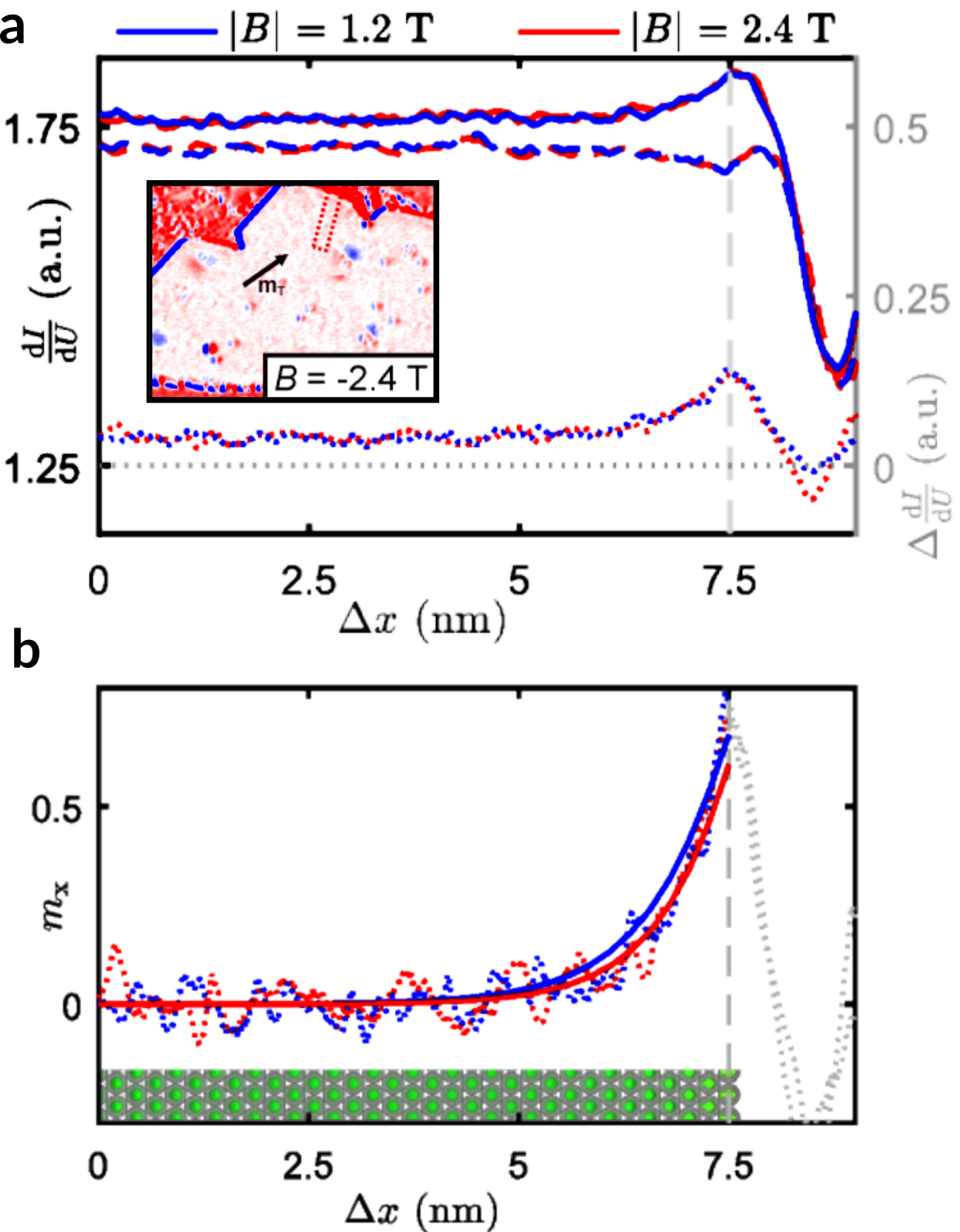}

    \caption{Investigation of the edge tilt in a Pd/Fe
    island. (a) Line profiles of $dI/dU$ near the
    boundary of a Pd/Fe island, see dotted red rectangle in the STM image,
    measured at positive (solid lines) and
    negative (dashed lines) magnetic fields (see Supplemental Material~\cite{suppmat}).
    The tip magnetization direction $m_{T}$ is dominantly
    in-plane and was obtained from
    2-dimensional fits to the skyrmions (see Supplemental Material~\cite{suppmat}).
    Dotted lines refer to the respective difference between the line
    profiles taken at positive and negative fields (scale on the right
    axis). (b) The experimentally derived in-plane $m_{x}$ components at
    the boundary for the two magnetic field values are shown as dotted lines
    together with the ones from the simulations as solid lines.}

\label{fig:Romming-field-sweep-line-profiles}
\end{figure}

The results of relaxing a particular experimentally found magnetic state at the
applied magnetic field are shown in the right columns of
Fig.~\ref{fig:Romming-field-sweep}. In general, the simulations are highly
accurate in reproducing the magnetic configurations from the experiments at the
corresponding applied fields. In particular, from
Fig.~\ref{fig:Romming-field-sweep}(q) the period of the spiral at zero field
of~(a) is estimated as 6.2~nm, in agreement with the characteristic helical
length period.  Discrepancies arise for the
magnetic states shown in Fig.~\ref{fig:Romming-field-sweep}(d), (e) and~(n). In
these cases, different factors that are not considered in the theoretical
model, such as the presence of defects, might be contributing to the
stabilization of the experimentally observed $360^\circ$ domain walls and
branches.

The topological charge $Q$ of each of the simulated spin textures in the Pd/Fe
island can be calculated according to the method presented in
Refs.~\onlinecite{Berg1981,Yin2016,Cortes2017}. This quantity
is a mathematical measure of the number of times that the spin orientations,
on a two dimensional plane, wrap a unit
sphere. It is useful, for example, for quantifying the number of skyrmions in a
system as the spin configuration of each skyrmion in an infinite sample
contributes with $Q=1$. For magnetic states in a confined geometry, such as
this Pd/Fe island, the topological charge is usually not an integer due to the
edge tilt of the spin texture at the boundary of the structure.

Figure~\ref{fig:Romming-field-sweep-line-profiles}(a) shows the line profiles
of the differential conductance near the edge of a larger Pd/Fe island at
four different applied magnetic fields, using a bias voltage of $U=+20$~mV. The
tip magnetization axis $m_{T}$ was derived from the appearance of the skyrmions
in this measurement and was found to be dominantly in the surface plane (see
inset to Fig.~\ref{fig:Romming-field-sweep-line-profiles}(a) and Sec.~S3
in the Supplemental Material~\cite{suppmat}). The in-plane component of the sample
magnetization can be calculated from the data presented in (a) and is plotted
as dotted lines for the two different absolute applied magnetic fields in (b).
The solid lines in Fig.~\ref{fig:Romming-field-sweep-line-profiles}(b)
represent the $m_x$-component of the respective simulated data. Although the
simulations do not include a possible change of the magnetic material close to
the edge due to changes of the electronic structure, the agreement with the
experimental profiles is remarkable. As evident from the experimental profiles
and the calculated magnetization component $m_x$, the DMI indeed leads to a
considerable tilting of spins at open boundaries of up to $\gtrapprox30^\circ$
with respect to the surface normal. The negligible difference between the
profiles at $B=1.2\,\text{T}$ and $B=2.4\,\text{T}$ shows the comparably small
influence of the external magnetic field on the edge tilt.

Comparison between Fig.~\ref{fig:Romming-field-sweep}(c) and~(d) demonstrates
that a $360^\circ$ domain wall does not have a topological charge. Because the
absolute value of the topological charge of a skyrmion is~1, and depending on
the applied magnetic field, we can conclude that for the island shown here the
edge tilt contributes to the total topological charge in a small range of $0.13
- 0.09$, which confirms the weak influence of the field. Analyzing the unusual
magnetic configurations found in the experiments,
Fig.~\ref{fig:Romming-field-sweep}(l), (m), we find that they have a topological
charge with the same sign as the external magnetic field, in contrast to the
typical skyrmion states (e.g.~(f)) where the topological charge and
magnetic field have opposite signs.


\begin{figure}[t!]
    \includegraphics[width=\columnwidth]{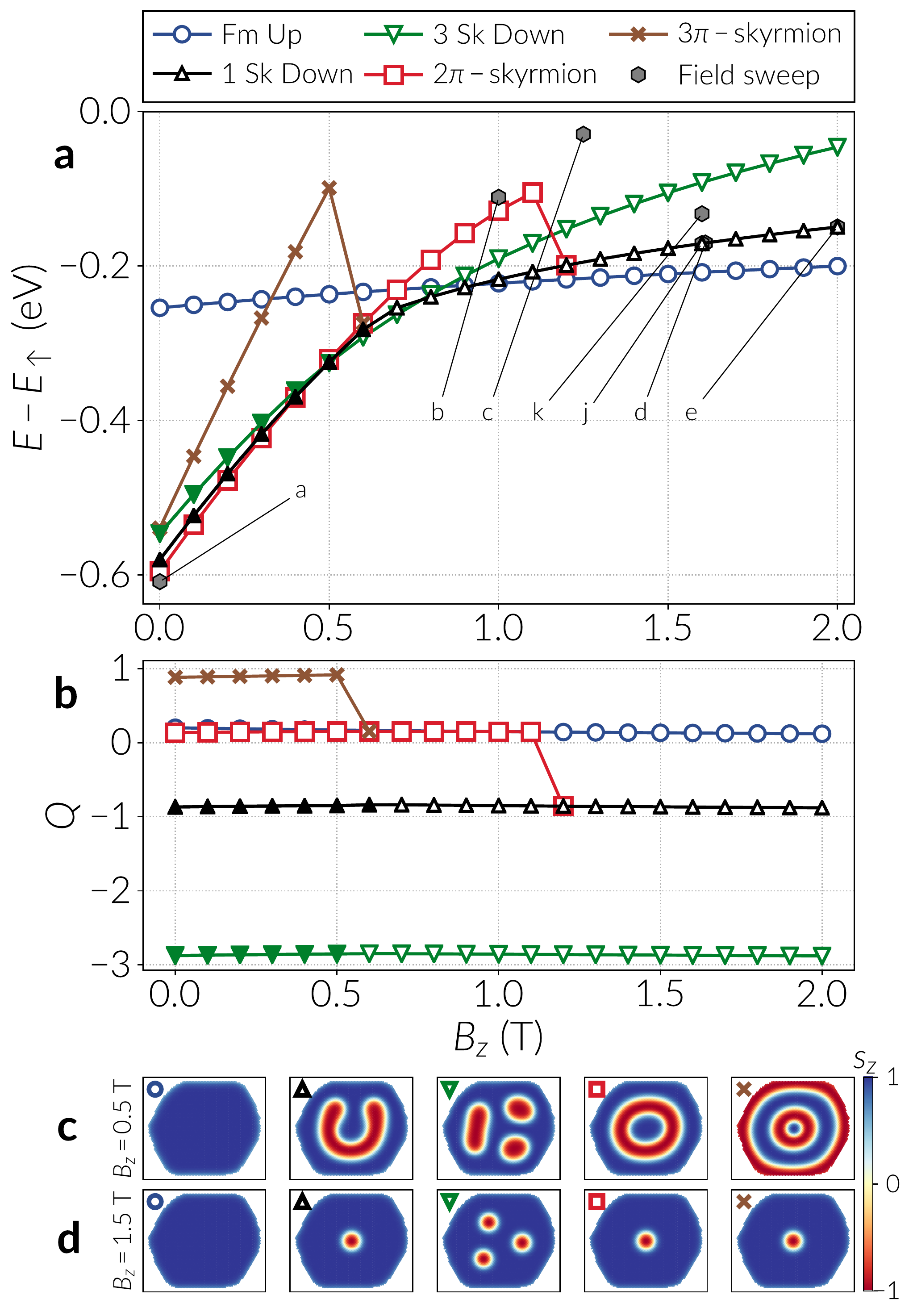}

    \caption{Total magnetic energy and topological charge of five different
        magnetic configurations: ferromagnetic ordering, (FM up), a single
        skyrmion (1 Sk Down), three skyrmions (3 Sk Down), a target state or
        $2\pi$-skyrmion, and a $3\pi$-skyrmion. Snapshots of these states under a
        field of $B_{z}=0.5\,\text{T}$ and $B_{z}=1.5\,\text{T}$ are shown at
        the two bottom rows of the figure, where the color scale refers to the
        out-of-plane spin component $s_{z}$.  Every plot is shown as a function
        of the applied magnetic field, which points in the $z$ direction of the
        sample. (a) Total energy computed with respect to a fully saturated state
        (denoted by $\uparrow$) at corresponding magnetic field strengths.
        (b) Topological charge. (c)-(d) Snapshots of the configurations at two
        different field values indicated at the left of every row.
    }

    \label{fig:island-energies-Q}
\end{figure}

\subsection{Simulations of magnetic states in the island}
\label{sec:simulations-islands}

To compare the experimentally observed magnetic states with other possible
equilibrium states we analyze the energy and the topological charge of different
simulated magnetic structures. For these simulations we define five different
initial states and relax them using the Landau-Lifshitz-Gilbert equation at
different magnetic fields, see Fig.~\ref{fig:island-energies-Q}: a uniform
ordering ({\large$\bullet$}) with edge tilt, a single skyrmion
($\blacktriangle$), three skyrmions
(\rotatebox[origin=c]{180}{$\blacktriangle$}), a target state ($\blacksquare$),
or $2\pi$-skyrmion, and a $3\pi$-skyrmion ($\bm{\times}$), which is a target
state with an extra rotation of the spins. At the bottom of
Fig.~\ref{fig:island-energies-Q} we show relaxed states of the five
configurations at applied fields of $B_{z}=0.5\,\text{T}$ and $1.5\,\text{T}$.
They do not look the same for both fields because lower energy states can be
found during relaxation or some of the configurations are unstable in specific
ranges of the applied magnetic field. Note that at weak magnetic fields the
skyrmions tend to occupy a larger area and are deformed because of the
inter-skyrmion force and the boundary repulsion.

The energies of the magnetic states (Fig.~\ref{fig:island-energies-Q}(a)) are
calculated with respect to the energy of a completely uniform configuration
where spins point in the $+z$ direction and that we denote by $\uparrow$.
Therefore, the energy curve labelled as ``Fm Up'' resembles the energy gain of the
edge tilt. The ground state of the system for magnetic fields close to zero is
likely to be a spiral with multiple rotations, since after analyzing the energy
of different variations of spirals obtained from the simulations we found that
they usually have lower energy than the five configurations shown. We do not
show the spirals here since they are not easily classifiable. With increasing
fields the Zeeman interaction becomes predominant and these spirals are not
energetically favored anymore.

We can distinguish some of the configurations by their topological charge,
which we show in Fig.~\ref{fig:island-energies-Q}(b). These values are not
integers due to the tilting of the spins at the boundaries. For domains
magnetized out-of-plane and enclosed by a $180^\circ$ domain wall, such as a
skyrmion, there is associated a topological charge of one. We find that the FM
state, the single skyrmion, and three skyrmions do not change their topological
number in the entire magnetic field range, whereas the $2\pi$-skyrmion and the
$3\pi$-skyrmion have a smaller stability range and decay into other magnetic
states: the $3\pi$-skyrmion state changes at $B_{z}=0.6\,\text{T}$ to a
$2\pi$-skyrmion state, and the $2\pi$-skyrmion state changes at
$B_{z}=1.2\,\text{T}$ to a single skyrmion state.


At weak fields, below 0.7~T, a single skyrmion becomes a worm-like domain, as
shown in the second image from the left in Fig.~\ref{fig:island-energies-Q}(c).
This transformation is known in the literature as strip-out or elliptic
instability.~\cite{Bogdanov1994a,Leonov2016} We mark the worm domains by filled
data symbols in Fig.~\ref{fig:island-energies-Q}(a) and~(b) (i.e.\ a filled
black triangle, instead of a hollow black triangle). A transition of such a
worm-like domain into a circular domain exists for $B_{z}\geq0.5\,\text{T}$ and
$B_{z}\leq0.7\,\text{T}$ (Fig.~S5 in the Supplemental Material~\cite{suppmat} shows both
configurations). This is confirmed by analyzing the energy of the magnetic
interactions involved, which are shown in Sec.~S6 of the Supplemental Material~\cite{suppmat}.
The worm domain has lower energy than a circular domain at weak fields because
of the DMI energy, which dominates over the Zeeman energy, and favors larger
areas of non-collinear spin ordering. Similar horseshoe-like states have been
observed previously in disk samples with DMI~\cite{Beg2015,Ho2019} for
sufficiently large system sizes. As the magnetic field increases, both the DMI
and Zeeman energies of a worm domain increase up to a point where the structure
is not energetically stable anymore and a transition into the circular domain
occurs, which sets the skyrmion strip-out field. Here, the DMI energy
abruptly increases but is followed by a steep
decrease of the energies of the three other interactions. In simulations, the
worm domain is only found by relaxing the system with a very strict (i.e.\
small) numerical tolerance of the algorithm (see simulation details in
Supplemental Material, Sec.~S7~\cite{suppmat}). For weaker (i.e.\ numerically greater)
tolerances we observe that the skyrmion relaxes to a circular bubble-like
domain.

We see in Fig.~\ref{fig:island-energies-Q}(a) that close to a field of
$B_{z}=0.8\,\text{T}$ a single skyrmion (which is observed above
$B_{z}=0.5\,\text{T}$) has lower energy than three skyrmions in the sample,
partially due to the Zeeman energy which is smaller for a larger ferromagnetic
environment in the direction of the applied field. Furthermore, above a field
of $B_{z}=0.8\,\text{T}$ a single skyrmion and three skyrmions start having
larger energy than the uniform configuration. In the case of three skyrmions,
it is still possible to observe worm domains, however their presence depends on
the available space in the island. This can be seen from
Fig.~\ref{fig:island-energies-Q}(c), where only one of the skyrmions is deformed
into an elongated domain at weak fields. Three skyrmion states with a worm
domain are also indicated with filled data points in
Fig.~\ref{fig:island-energies-Q}(a) and~(b). The presence of a small elongated
domain also affects the energy of the system, where the transition occurs at
fields from 0.5~T up to 0.6~T, but not as abruptly as in the case of a single
skyrmion because of the smaller skyrmion size. As indicated in Fig.~S4 of the
Supplemental Material,~\cite{suppmat} although the DMI energy behaves similarly as in the
single skyrmion case, the Zeeman energy does not change significantly when the
small worm domain turns into a circular skyrmion, hence other effects such as
the skyrmion-skyrmion interaction, the island confinement and the repulsion
from the boundary, become more important in hindering the formation of spiral
domains.


Two other physically interesting states are the $2\pi$ and $3\pi$ skyrmions.
For the latter there is still no experimental evidence in the literature. In
the hexagonal islands we obtained these multiple-rotation states defining the
direction of their inner core in the $+z$ direction, which follows the applied
field orientation. 

Referring to Fig.~\ref{fig:island-energies-Q}(a), the $2\pi$ and $3\pi$-skyrmion
states are only stable in a small range of field magnitudes. Specifically,
starting from zero applied field, the $2\pi$-skyrmion is observed up to a field
of $B_{z}=1.1\,\text{T}$ and then the system relaxes to a skyrmion for larger
field magnitudes, which is evident from the topological charge in
Fig.~\ref{fig:island-energies-Q}(b). For the case of the $3\pi$-skyrmion, it is
visible even in a smaller range, only up to $B_{z}=0.5\,\text{T}$,
and afterwards it relaxes to the target and skyrmion states (this value
slightly depends on the tolerance of the relaxation, weak tolerance allows to
stabilize a metastable target state in a small range above 0.5~T). It is
noteworthy that, for low fields, the multiple spin rotations of the
$2\pi$-skyrmion together with a large enough ferromagnetic background make this
configuration the lowest energy state between the five states we are analyzing.
The uniformly oriented background of spins is important because it decreases
the overall exchange energy. We can explain the higher energy of the
$3\pi$-skyrmion compared to the $2\pi$-skyrmion at fields below $0.5\,\text{T}$
(where the $3\pi$-skyrmion exists), by referring to the exchange and DMI
energies (see Supplemental Fig.~S4(c) and~(d)~\cite{suppmat}), which are the main energy
contributions. The DMI energy of the $3\pi$-skyrmion is significantly smaller
than the $2\pi$-skyrmion, which is caused by the extra rotation of the magnetic
moments, however the effect of both the exchange and Zeeman interactions is
sufficiently large to make the total energy of the $3\pi$-skyrmion larger. This
phenomenon is enhanced by the anisotropy energy (see Fig.~S4(e) in the
Supplemental Material~\cite{suppmat}).

In Fig.~\ref{fig:island-energies-Q}(a) we observe that target states in
Pd/Fe/Ir(111) islands cannot be stabilized above a field of
$B_{z}=1.1\,\text{T}$ since the system relaxes to an isolated skyrmion. One
possible reason for this instability is that the energy barrier separating
these two configurations is reduced as the magnetic field increases, thus a
target state would decay above a critical field. We confirmed this hypothesis
by performing a stability simulation between a target state and a skyrmion by
means of the GNEBM.~\cite{Bessarab2015,Cortes2017} We show these results in
Sec.~S8 of the Supplemental Material,~\cite{suppmat} where we find that the critical field
where the barrier goes to zero, lies between $B_{z}=1.1\,\text{T}$ and~1.2~T.


To compare the energy of the experimentally observed configurations with the
energy of the skyrmionic configurations, we show the energies of the
corresponding simulated magnetic states (Fig.~\ref{fig:Romming-field-sweep}) in
Fig.~\ref{fig:island-energies-Q}(a) as hexagonal markers and annotated with their
corresponding letter from Fig.~\ref{fig:Romming-field-sweep}. At zero field we
observe that the spin spiral (Fig.~\ref{fig:Romming-field-sweep}(a)) has the
lowest energy, slightly below the energy of the target state. At larger fields
a skyrmion coexisting with a $360^{\circ}$ wall
(Fig.~\ref{fig:Romming-field-sweep}(b) and~(c)) has a higher energy compared to the
other states, the single skyrmion states confirm our simulation results by
lying on the corresponding curve, and, as expected, the double skyrmion energy
lies between the single and triple skyrmion curves at 1.6~T.


\begin{figure*}[t!]
    \includegraphics[width=0.75\textwidth]{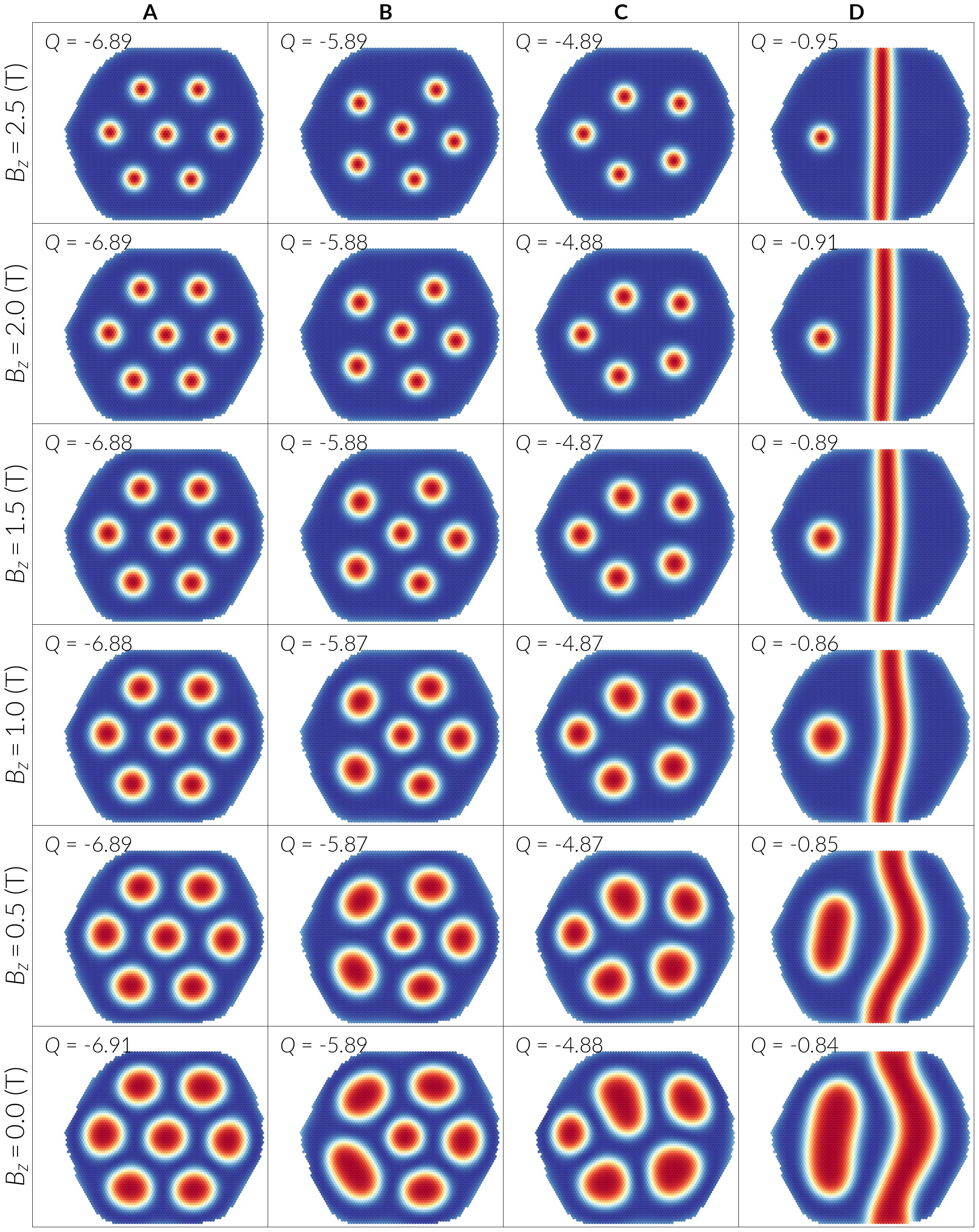}

    \caption{Field sweep of magnetic orderings in a hexagonal island. Snapshots
    of magnetic configurations obtained after relaxation during a field sweep,
    as in a hysteresis process, with a starting field of
    $B_{z}=2.5\,\text{T}$. Columns indicate the field variation process for
    different initial states. The field sweep was computed in steps of
    $0.1\,\text{T}$, with this figure only showing the resulting orderings in
    steps of $0.5\,\text{T}$. The field magnitude is depicted at the left side
    of every row, where all the snapshots are shown at the same field value.
    The top left number at every snapshot plot refers to the topological charge
    of the system. The color scale indicates the out-of-plane component of
    spins which is the color scale of the snapshots in
    Fig.~\ref{fig:island-energies-Q}.}

    \label{fig:sims-field-sweep-I}
\end{figure*}

\begin{figure*}[t!]
    \includegraphics[width=0.75\textwidth]{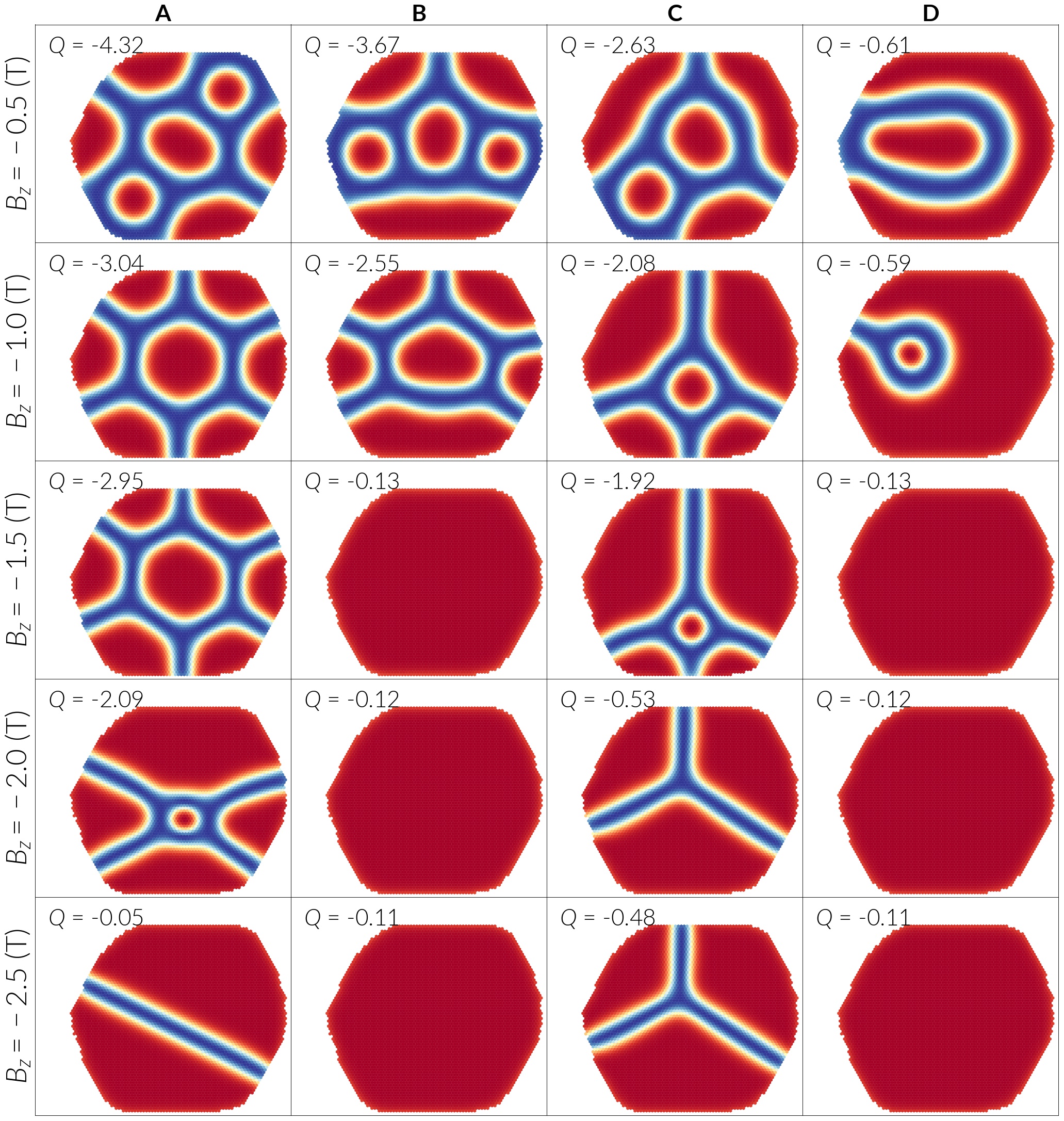}

    \caption{Field sweep of magnetic orderings in a hexagonal
    island. Continuation of the field sweep process explained in
    Fig.~\ref{fig:sims-field-sweep-I}. Field magnitudes for
    this case are negative and smaller than zero field, thus the field is
    oriented out-of-plane in the $-z$ direction.}

    \label{fig:sims-field-sweep-II}
\end{figure*}

\subsection{Simulations of a magnetic field sweep}
\label{sec:sim-field-sweep}

With the aim of devising a reproducible method to obtain the magnetic
configurations observed during the field sweep experiment, we replicated this
process in our simulations by starting with different initial configurations at
strong fields.  An adequate choice of initial states is not straightforward
since it is often not possible to know every possible equilibrium state. In
particular, spin spirals are difficult to specify in simulations because they
can manifest as branches, i.e.\ long domains oriented in the $+z$ or $-z$
direction, that bend and extend in any possible direction in the island. Hence,
we selected five magnetic states based on a different number of skyrmions and
the configuration from experiment of Fig.~\ref{fig:Romming-field-sweep}(b) where
a skyrmion and a $360^{\circ}$ wall coexist next to each other. We do not show
results starting from the spiral state of Fig.~\ref{fig:Romming-field-sweep}(a)
since at sufficiently large fields it turns into a single skyrmion before
saturating, and reversing the field it is not possible to obtain the spiral
states from the experiments. The reason is that our simulations are performed
at zero temperature and we do not consider the tunnel current from the STM tip,
which can excite some of the observed states. Nevertheless, simulations show
that these spiral configurations are still accessible when choosing an
appropriate initial condition, as shown by the simulation of different states
(Fig.~\ref{fig:Romming-field-sweep}) of the field sweep experiment discussed in
Sec.~\ref{sec:exp-field-sweep}.

\begin{figure*}[t!]
    \includegraphics[width=\textwidth]{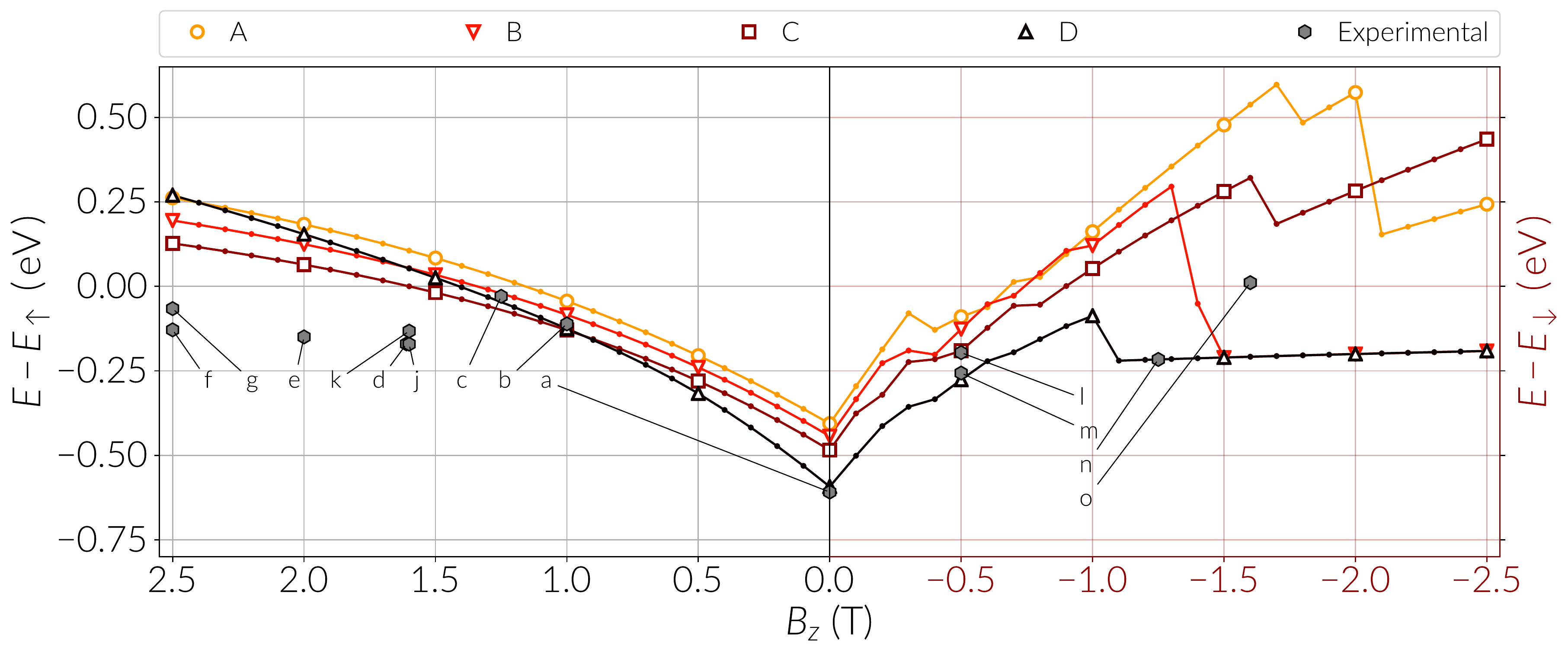}

    \caption{Energy of the configurations in the field sweep process. At
        positive fields (left panel) the energy is computed relative to the energy of the fully saturated state in the $z$-direction, $E_{\uparrow}$, whereas at negative fields (right panel) the reference is the saturated state in the opposite direction $E_{\downarrow}$.}

\label{fig:sims-field-sweep-energy}
\end{figure*}

We set our starting point in the field sweep by relaxing these configurations
at a magnetic field of $B_{z}=2.5\,\text{T}$, which are the states in the first
row of Fig.~\ref{fig:sims-field-sweep-I}. We characterize them by their
topological number $Q$, whose absolute value effectively indicates the number
of skyrmions in every snapshot. For a spin spiral $Q\approx 0$ because it only
has spins rotating in a single direction. We then decrease the magnetic field
in steps of $-100\,\text{mT}$ for every initial state and relax them again,
repeating the process down to a field of $B_{z}=-2.5\,\text{T}$ and registering
the resulting configurations under equilibrium for every step. As a result, we
obtain the evolution of every initial state given by the columns of
Fig.~\ref{fig:sims-field-sweep-I} and \ref{fig:sims-field-sweep-II}, where we
only show snapshots of the relaxed states in steps of $500\,\text{mT}$. We
observe that as we reach zero field, skyrmions expand and occupy a larger area
in the sample. When there is a large number of them, as seen in the case of
column~A of Fig.~\ref{fig:sims-field-sweep-I} at zero field, these
configurations become more symmetrical in shape because of the restricted
available space given by the ferromagnetic background. The topological number
of the different magnetic configurations changes only slightly during the
magnetic field sweep to $0\,$T.

When the zero field is crossed, however, we observe interesting magnetic
configurations which critically depend on the initial state, see
Fig.~\ref{fig:sims-field-sweep-II}. When we start with different numbers of
skyrmions we notice that domains oriented in the $-z$ direction appear
surrounded by elongated domains pointing in the $+z$ direction in the form of
branches whose ends tend to finish perpendicular to the hexagon sides rather
than the corners, reminiscent of the magnetic states found experimentally.
Indeed, they also show a topological number with the same sign as the applied
field, as observed in the experimental data. A plausible explanation for the
spirals avoiding the island corners is that, firstly, compared to a straight
edge, a corner covers a larger area (about an extra 15\% surface if we assume a
perfect hexagon) and thus surface effects are stronger and involve an energy
cost. Secondly, a spiral sitting at the corner is not optimum as the cycloid
needs to rotate $360^\circ$ and the corner reduces the available space to do
this. Finally, the DMI imposes a boundary condition where spins are tilted and
orient perpendicular to the island edge surface, making it energetically
inefficient to form a spiral at the corner rather than in a smoother surface.

Whereas the phenomenological counting of the topological charge is rather
trivial for the magnetic states shown in Fig.~\ref{fig:sims-field-sweep-I}
where each skyrmion contributes a charge of one, the topological charge of the
states in Fig.~\ref{fig:sims-field-sweep-II} appears less obvious at first
sight. However, by carefully analyzing the magnetic states, we propose a
phenomenological model to also estimate the topological charge of such unusual
magnetic states. The observed enclosed domains carry a topological charge of
$+1$ since their surrounding domain wall are equivalent to the skyrmion
boundary. While some of the skyrmionic structures are partially destroyed at
the boundary of the island, the total number of domains in the $-z$ direction
is the same as the initial number of skyrmions, which is evident from columns
A-C of Fig.~\ref{fig:sims-field-sweep-II} at fields of -0.5 and
$-1.0\,\text{T}$.  As expected, these incomplete skyrmionic domains do not have
an integer topological charge and it is possible to roughly estimate this
magnitude depending on the domain shape. Incomplete skyrmionic domains sitting
in a corner at the boundary of the hexagonal island, which forms a $2\pi/3$
angle, can be identified with $|Q|\approx1/3$ because they make a third of a
full skyrmionic texture. However, when the domains are more elongated their
surrounding domain walls are likely to give a small $|Q|$ value. Depending on
the flatness of the domain walls, some of these domains can be associated with
a charge of $|Q|\approx1/6$.

For example, referring to column~A of Fig.~\ref{fig:sims-field-sweep-II}, at
$B_{z}=-0.5\,\text{T}$ we have three skyrmionic domains which each contribute
$|Q|=1$ and four incomplete domains that can be identified with
$|Q|\approx1/3$, approximately summing up the total charge $|Q|$ of $4.32$.
Furthermore, if we now refer to the state of column~C at the same field, we
observe elongated boundary domains at the upper sides of the sample with a more
straight profile of their surrounding domain walls, but not completely
flat. Hence, by only considering the domain at the bottom right of the island
as a third of a skyrmion, we obtain a total charge of
$|Q|\approx2+1/3+2\times1/6=2.66$, being the actual total charge $|Q|=2.63$.
Similarly, the state from column~B at $-1.0\,\text{T}$ is estimated with
$|Q|\approx2.5$. In contrast, the flat profile of the domain wall from the
bottom domain of the state of column~B, at $B_{z}=-0.5\,\text{T}$, makes the
contribution of this domain to be zero and hence we estimate a $|Q|$ of 3.66
for the whole configuration in the island. 

Another way to estimate the $|Q|$ magnitudes, is associating the triple
junctions from the $360^\circ$ walls with a charge of $|Q|\approx0.5$.  Hence,
discarding the boundary contribution, states from the row at $-1.0\,\text{T}$
in Fig.~\ref{fig:sims-field-sweep-II} agree well with this estimation. The
slight difference from the spin tilting at the boundaries can contribute up to
$|Q|\approx0.13$, as seen in column~D at $-1.5\,\text{T}$ for a uniform
ordering in the $-z$ direction.

\begin{figure*}[t!]
    \includegraphics[width=\textwidth]{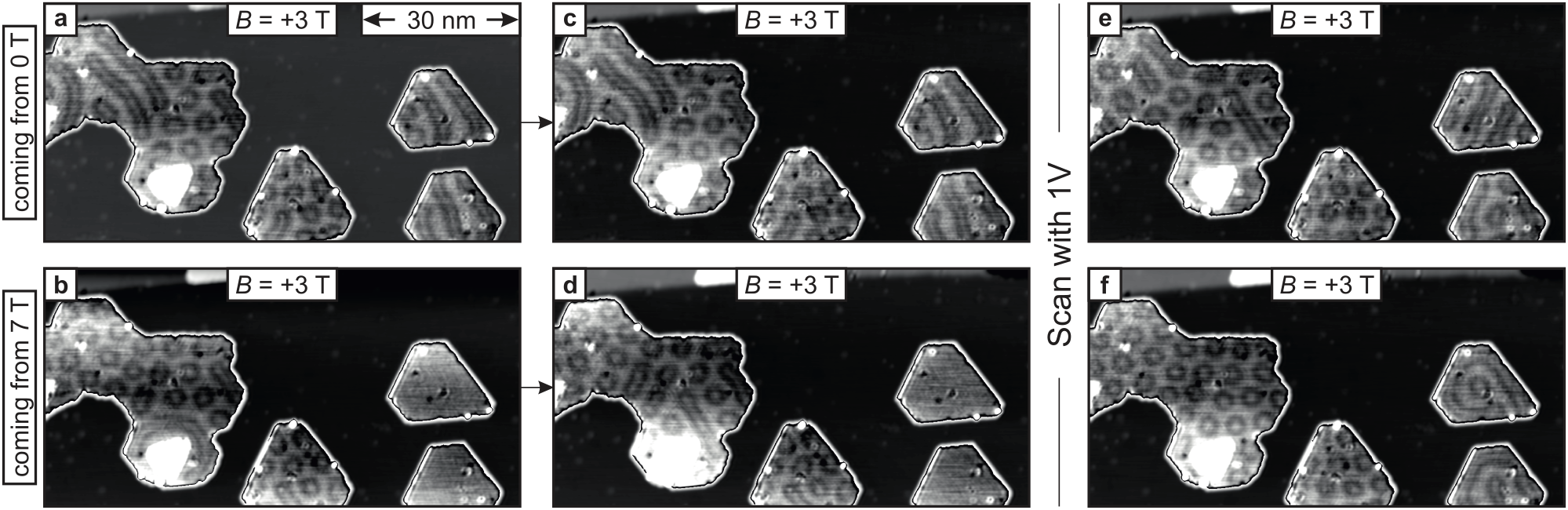}

    \caption{Magnetic states of Pd islands on an extended Pd/Fe film on Ir(111)
        imaged at $B=+3$\,T. STM constant-current images with the NCMR contrast
        on the islands adjusted locally to $\pm 10\,\text{pm}$ after applying a
        Gau\ss-filter with $\sigma =1.2\,\text{\AA}$ in the slow scan
        direction, measurement parameters for all images: $U = +50\,\text{mV}$,
        $I = 1\,\text{nA}$, $T = 4.2\,\text{K}$, Cr-bulk tip.  (a) Image
        obtained after a magnetic field sweep from $B=0\,\text{T}$ up to +3~T.
        (b) Image obtained after a magnetic field sweep from $B=+7\,\text{T}$
        down to +3~T.  Images (c) and~(d) were taken after (a) and~(b),
        respectively, using the same bias voltage and tunnel current, showing
        only minor changes.  The areas (c) and~(d) were subsequently scanned
        with $U=+1\,\text{V}$ and $I=3\,\text{nA}$ (these STM images are not
        shown). Finally, images (e) and~(f) were obtained with the initial
        parameters $U=+50\,\text{mV}$ and $I=1\,\text{nA}$, demonstrating the
        changes in the magnetic state due to the larger bias voltage and tunnel
        current.  }

\label{fig:PdPdfeIr-exp}
\end{figure*}

To compare the energy of the states from the experimental field sweep with the
configurations obtained from the simulations, we plot the energies of all the
configurations as a function of the applied field in
Fig.~\ref{fig:sims-field-sweep-energy}, where the field axis follows the order
of the sweep process from left to right. Since the field orientation changes in
the field sweep process, in the positive field region, shown at the left panel
of Fig.~\ref{fig:sims-field-sweep-energy}, the energies are computed with
respect to the fully saturated state energy in the $+z$ direction,
$E_{\uparrow}$.  Similarly, at negative fields, we show the energy as a
function of the saturated state in the opposite direction, $E_{\downarrow}$, in
the right side of Fig.~\ref{fig:sims-field-sweep-energy}. Energies of the
states from the experiment are shown as annotated data points with the labels
according to Fig.~\ref{fig:Romming-field-sweep}. The four simulated field sweep
processes are distinguished in the top legend by the column letter of
Fig.~\ref{fig:sims-field-sweep-I} and~\ref{fig:sims-field-sweep-II}. At the
center of Fig.~\ref{fig:sims-field-sweep-energy} we can see that the energy of
the zero field spin spiral (data point~a) is closest to the zero field state of
curve~D. This is consistent with the experimental results, where the spin
spiral evolves into a skyrmion next to a $360^{\circ}$ wall as in
Fig.~\ref{fig:Romming-field-sweep}(b) and~(c), and the energies of these
configurations are close to the energies given by curve~D of
Fig.~\ref{fig:sims-field-sweep-energy}.
We can notice a somewhat larger energy in the experimental configurations
because the orientation of the spin spiral is different in the experiment. This
indicates that the spin spiral prefers to orient perpendicular to a longer or
smoother edge of the island. At higher fields the experimental states, which
are mainly isolated skyrmions, have a significant lower energy because the
simulated configurations have larger regions of spins opposite to the field,
which translates into a larger Zeeman energy cost. At negative fields, similar
energies are found from the complex spiral configurations of
Fig.~\ref{fig:Romming-field-sweep}(l) and~(m), with the energies of the states in
curves~C and~D at -0.5~T. In particular, the configuration from curve~D
(Fig.~\ref{fig:sims-field-sweep-II} at -0.5~T) resembles
Fig.~\ref{fig:Romming-field-sweep}m but with the left side of the domain
shifted to the right side.

We find that the unusual magnetic states (right plot of
Fig.~\ref{fig:sims-field-sweep-energy}) have higher energy than the skyrmion
configurations (left plot of Fig.~\ref{fig:sims-field-sweep-energy}), i.e.\
they are metastable states, which are accessible only at sufficiently low
temperatures.  In particular, such magnetic field sweeps at low temperature
provide access to novel exotic spin textures with non-trivial topological
charge. However, while the target state is not a high energy state, it was not
observed during neither the experimental nor the simulated magnetic field sweep
experiments. Analyzing the resulting magnetic states one can speculate that the
edge tilt and the resulting preference of $360^\circ$ domain walls to be
perpendicular to the edges, hinders the formation of a target state.


\section{Magnetic states in Pd$_2$/Fe/Ir(111) islands}
\label{sec:experiment-PdPdFeIr}

From simulations we will show that, in general, confinement can stabilize
target skyrmions at weak fields (roughly below $0.9~\text{T}$) in comparison to
extended films. However, both the experiments and field sweep simulations for a
nanometer-sized island of Pd/Fe on Ir(111) have not been able to produce a
target state and analyzing the observed magnetic states suggests that the edge
tilt at the boundary might be responsible. 


We now turn to a different system, namely a Pd island on a Pd/Fe extended film.
Our STM measurements, see Fig.~\ref{fig:PdPdfeIr-exp}, show that the Pd$_2$/Fe
island also exhibits a spin spiral state which is modified upon the application
of external magnetic fields. The observed modulations on the islands are of
magnetic origin, to be precise we attribute them to the non-collinear
magnetoresistance effect (NCMR),~\cite{Hanneken2015} which is sensitive to
changes of the local spin texture: in this case a higher signal (brighter) is
observed for smaller angles between nearest neighbor spins, and a lower
(darker) signal indicates larger angles between nearest neighbor spins. This
means that the in-plane rotations of the spin spiral or the skyrmions appear
darker, leading to two dark lines for one spin spiral period and a ring-like
feature for each skyrmion (see also Supplemental Material Sec.~S2~\cite{suppmat}). 

The spin spiral wavelength is about $4.6\,\text{nm}$ and thus slightly shorter
than that of the surrounding Pd/Fe. Importantly, the transition fields to other
magnetic states are much higher in the Pd$_2$/Fe, resulting in a situation
where at about $3\,\text{T}$ the Pd$_2$/Fe islands still show complex magnetic
order but are surrounded by a fully field-polarized Pd/Fe film. This means that
in contrast to the effectively non-magnetic environment of a Pd/Fe island, as
in Fig.~\ref{fig:Romming-field-sweep}, the boundary condition in the case of a
Pd island on a Pd/Fe extended film under the applied magnetic field of 3~T is
ferromagnetic. 

Figures~\ref{fig:PdPdfeIr-exp}(a) and (b) show the same sample area at a
magnetic field of $B = +3\,\text{T}$, but with a different magnetic field sweep
history. Before the measurement shown in Fig.~\ref{fig:PdPdfeIr-exp}(a), the
magnetic field was swept up from the virgin state at $B = 0\,\text{T}$ (see
also Supplemental Material Sec.~S4~\cite{suppmat}). Figure~\ref{fig:PdPdfeIr-exp}(b), on the
contrary, is measured directly after sweeping down the magnetic field from $B =
+7\,\text{T}$. We observe a hysteretic behavior: during the up-sweep of the
external magnetic field we find remainders of the zero-field spin spiral state
(a), whereas in the down-sweep the two small islands on the right side of (b)
are still in the field-polarized state and the larger islands are dominantly in
the skyrmion lattice phase. In both cases small changes can be observed when
the area is imaged again, see (c) and (d). We attribute the changes to either a
delayed relaxation of the magnetic state towards its thermodynamic equilibrium
or a non-negligible influence of the tunnel current of $1\,\text{nA}$ with a
bias voltage of $U = +50\,\text{mV}$.

More significant changes are observed, after the area has been scanned with
tunnel parameters of $1\,\text{nA}$ and $U=+1\,\text{V}$, see (e) and (f). The
application of higher energy tunnel electrons is known to affect the magnetic
state of a sample,~\cite{Romming2013,Hsu2016} which is also evident here by a
comparison from Fig.~\ref{fig:PdPdfeIr-exp}(c) to~(e), and (d) to~(f). Now the
large island dominantly exhibits a skyrmion lattice in both cases. In the upper
right island the larger bias voltage has induced a change of the spin spiral
propagation direction from (c) to (e). In the lower right island in (c) and in
the case where the same two small islands on the right were in the
field-polarized state (d), the tunnel electrons induced target-like states, see
(e),(f). 

To interpret the target-like states of the small islands in (f) we can start
with the outer rim, which we assume to be dominantly parallel to the
surrounding field-polarized state of the Pd/Fe extended layer, i.e.\ the
boundary of the island is parallel to the external field. The first dark ring
marks the rotation of the spins through the in-plane magnetization to the
region where the spins are opposite to the applied field; note that in the
upper island this dark ring is not entirely within the island but appears to be
pinned to a defect at the island boundary. The second dark ring is another spin
rotation through in-plane, meaning that in the center the observed target
skyrmion is again parallel to the applied external magnetic field. Comparison
with the magnetic states observed in the Pd/Fe island of
Fig.~\ref{fig:Romming-field-sweep} with effectively open boundary conditions
suggests that the ferromagnetic surrounding is responsible for the generation
and stabilization of the target state in the Pd$_2$/Fe islands.

\begin{figure}[t!]
    \includegraphics[width=\columnwidth]{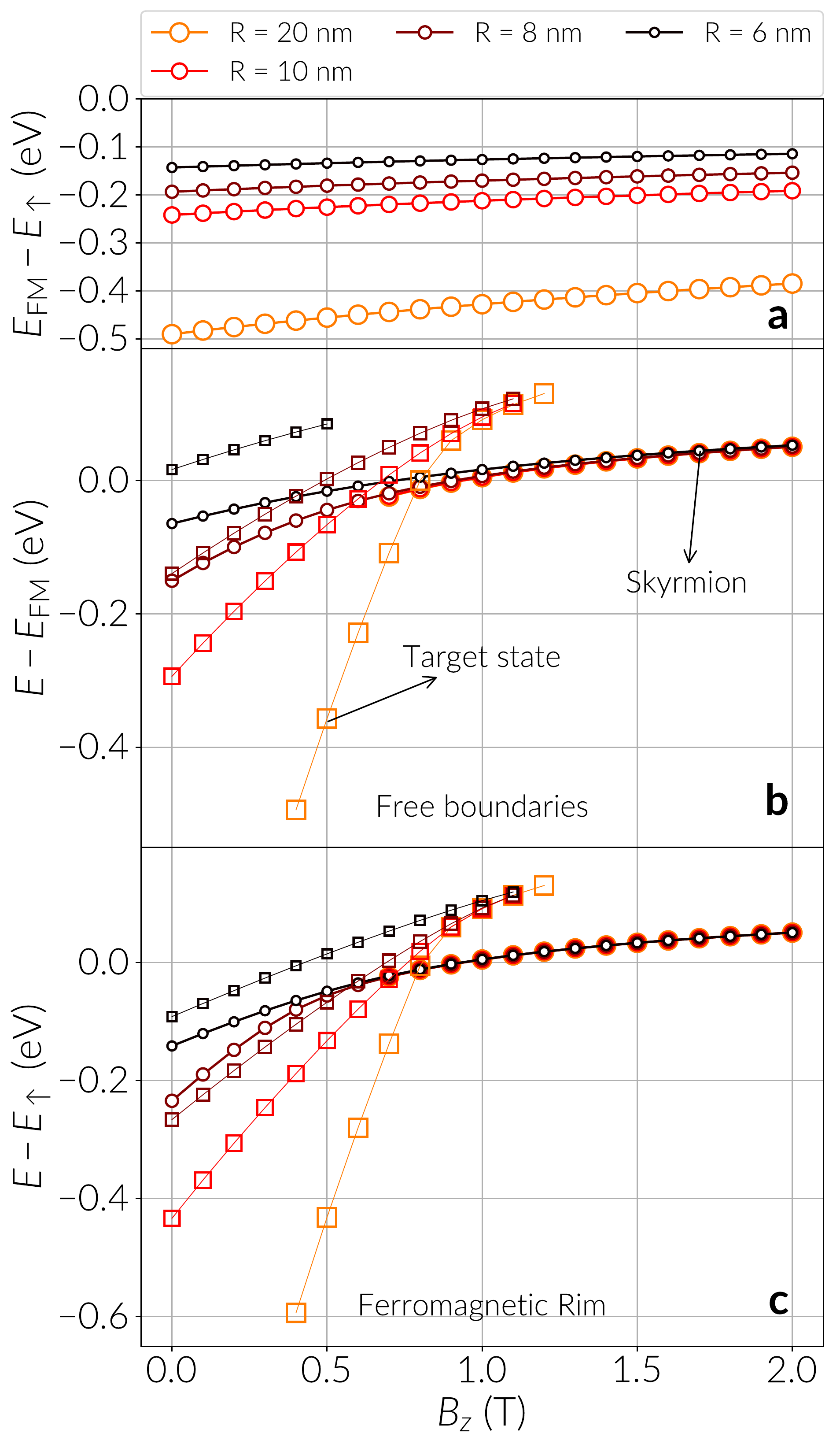}

    \caption{Energy of the ferromagnetic boundary, skyrmion and target state
        in Pd/Fe/Ir(111) hexagons, as a function of the applied field
        for different hexagon sizes and boundary conditions.
        The hexagon size is defined by its circumradius $R$. (a) The curve
        refers to the energy of the ferromagnetic state $E_{\text{FM}}$ with
        respect to the energy of the fully aligned configuration (in the
        $+z$ direction) $E_{\uparrow}$, as a function of the applied
        field. (b) Energy $E$ of the skyrmion (circles) and target states
        (squares) with respect to the energy of the ferromagnetic state, as a
        function of the applied field. In all configurations (skyrmion, target state and
        ferromagnetic state), spins are tilted at the boundary. (c) Energy of
        the skyrmion and the target state with respect to the fully uniform configuration in an
        island with a ferromagnetic rim (the ferromagnetic state in this case
        corresponds to the fully aligned configuration).  }

    \label{fig:hexagons-R-energy-skyrmion}
\end{figure}

\section{Skyrmion and target state stability analysis}
\label{sec:sim-hexagons}

To enhance the general understanding on the occurrence of skyrmions and target
states in the Pd/Fe and Pd$_2$/Fe islands we perform a systematic simulation
based study of hexagons, characterized by their circumradius $R$, where we vary
the hexagon size and the out-of-plane applied field $B_{z}$.  We simulate the
Pd$_2$/Fe islands using the same magnetic parameters as the Pd/Fe islands, but
changing the boundary conditions to a ferromagnetic rim, according to the
discussion in Sec.~\ref{sec:experiment-PdPdFeIr}. Although a more complete
model should account for the phase changes of the Pd/Fe surrounding a
Pd$_2$/Fe island at weak fields, a simplified model allows to analyze the
effects of modifying the edge condition. This picture is nonetheless accurate
enough at stronger fields, where it was shown that the island surrounding is
field-polarized.  The strategy we employ in the simulations is to start with
either a skyrmion or a target state and relax the system with the LLG equation
without the precession term, as in Sec.~\ref{sec:simulations-islands}.


\begin{figure}[t!]
    \includegraphics[width=\columnwidth]{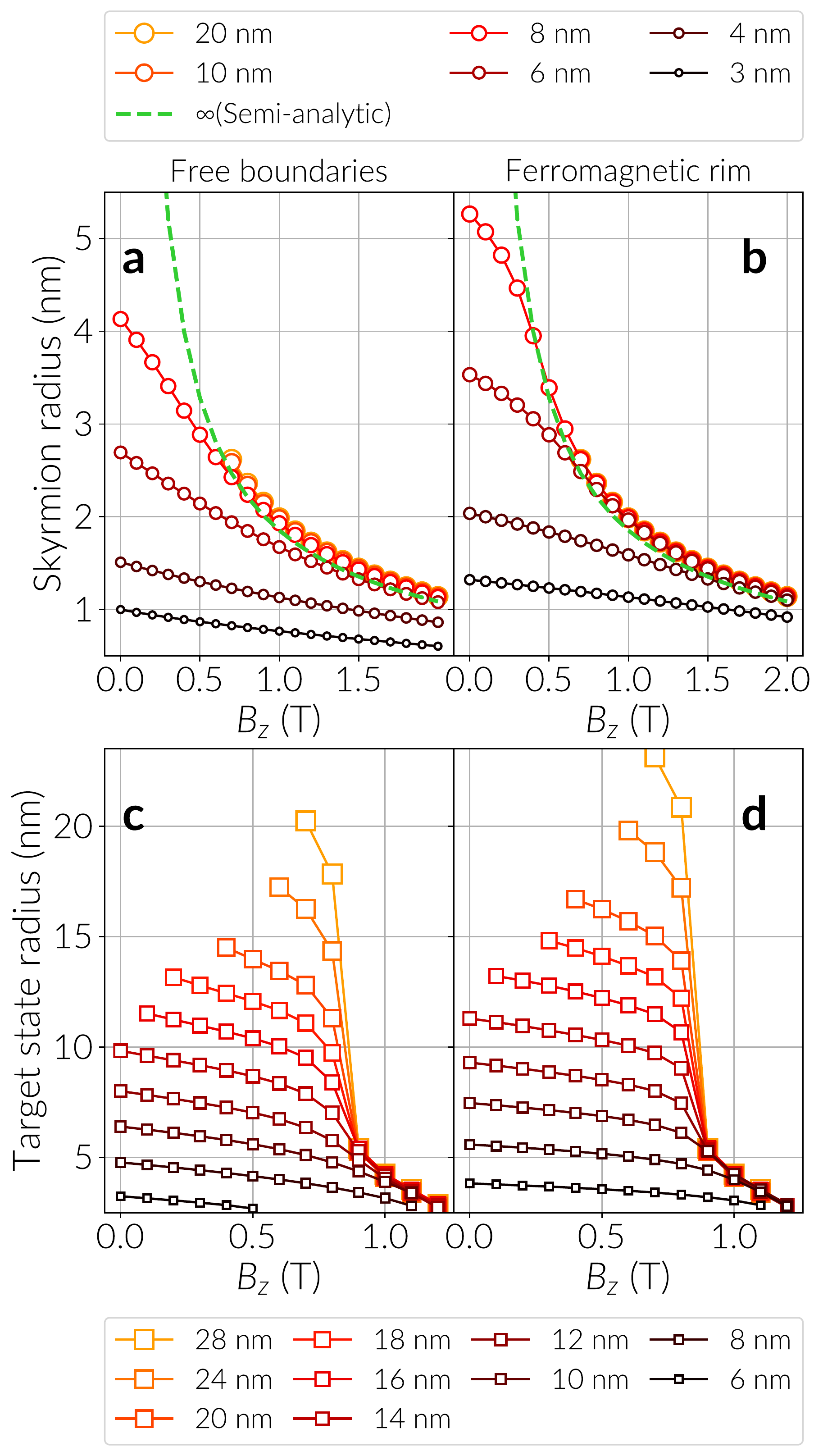}

    \caption{Skyrmion and target state size as a function of the magnetic field
        and hexagon size. Hexagons are measured by their circumradius $R$.
        Skyrmion sizes are computed for hexagons with free~(a) and
        ferromagnetic~(b) boundaries. No data points are shown for
        $R\geq10\,\text{nm}$ and weak fields since the system relaxes into a
        domain resembling an elongated skyrmion or into a multi-spiral state.
        A semi-analytical theoretical curve for skyrmion size as a function of
        applied field in an infinite film (no confinement),
        from Ref.~\onlinecite{Wang2018}, is shown and denoted by
        $\infty$.  Target state sizes in hexagons with free boundaries or a
        ferromagnetic rim are depicted in (c) and~(d), respectively.  No data
        points are shown for $R\leq6\,\text{nm}$ and sufficiently strong fields
        when using free boundaries because the system relaxes into a skyrmion.}

    \label{fig:sks-tgt-st-radius-vs-field}
\end{figure}

\subsection{Size dependence and boundary conditions}

We define islands of sizes from $R=4\,\text{nm}$ up to $20\,\text{nm}$ in steps
of $0.5\,\text{nm}$ and initialize three magnetic configurations: a skyrmion
profile of a radius about 1/4 of the hexagon circumradius, a target state
profile of about the size of the island and a ferromagnetic configuration. We
firstly compute the energy contribution from the tilted spins at the boundary
of the hexagons by subtracting the energy of the fully saturated configuration
(in the $+z$ direction) from the energy of the ferromagnetic state. The former
is denoted by $E_{\uparrow}$ and the latter by $E_{\text{FM}}$. We show this
result in Fig.~\ref{fig:hexagons-R-energy-skyrmion}(a) with the energy
differences as a function of the applied field. Results are shown for islands
from $R=6\,\text{nm}$ up to $20\,\text{nm}$ since target states are not stable
at $R=4\,\text{nm}$. We notice that the boundary energy gain with increasing
fields is more pronounced for larger islands.

In Fig.~\ref{fig:hexagons-R-energy-skyrmion}(b) we calculate the energy of the
skyrmion (circle markers) and target state (square markers) configurations with
respect to the ferromagnetic state in islands with free boundaries. We compare
this result with Fig.~\ref{fig:hexagons-R-energy-skyrmion}(c), where we depict
the energy of the two states with respect to the ferromagnetic configuration in
islands with a ferromagnetic rim. In this case of pinned-spin boundaries, the
ferromagnetic state corresponds to the fully saturated state.

From the plots we observe that when ferromagnetic boundaries are present
(Fig.~\ref{fig:hexagons-R-energy-skyrmion}(c)) the skyrmion has, in general,
lower energy than the ferromagnetic state for fields weaker than 1.0~T. In the
case of free boundaries (Fig.~\ref{fig:hexagons-R-energy-skyrmion}(b)), where the
boundary spins decrease the overall energy of the system, for $R=6\,\text{nm}$
there is a slight shift of the field value below which the skyrmion has lower
energy than the uniform state.  This effect is more pronounced for smaller $R$
(see Supplemental Material, Sec.~S10~\cite{suppmat}), where the skyrmion always has larger
energy. We also see from Figs.~\ref{fig:hexagons-R-energy-skyrmion}(b) and~(c) that
for $R\geq6\,\text{nm}$ and above $B_{z}=0.6\,\text{T}$, the energies of the
skyrmion for different $R$ start to get closer as the field increases. With
free boundaries a small energy difference is present, however with a
ferromagnetic rim this difference is unnoticeable, which means skyrmions are
not influenced by the spins at the boundary. In the case of free boundaries,
the influence of boundary spins vanish for sufficiently strong fields, where
the skyrmion size becomes sufficiently small. In
Figs.~\ref{fig:hexagons-R-energy-skyrmion}(b) and~(c) we do not show skyrmion
energies for $R\geq10\,\text{nm}$ and $B_{z}\leq0.6\,\text{T}$ because
skyrmions could not be stabilized and stripe or worm domains appear as the
final stable configuration. In this context, the DMI dominates over the Zeeman
interaction, and a larger island size means confinement effects on the
magnetic configuration are weaker.

Regarding target states, the critical field below which target states have
smaller energy than the ferromagnetic state, depends more critically (steeper
energy curves) on the hexagon size $R$ than in the skyrmion case, owing to the
larger number of non-collinear spins and thus, a larger influence of the DMI
energy. The range of fields where boundary spins do not have an influence is
substantially smaller than in the skyrmion case, which is only noticed for the
case of a ferromagnetic rim close to a field of $1\,\text{T}$. Furthermore,
target states are more affected by the available space in the island since for
free boundaries, at small hexagon sizes, $R\approx6\,\text{nm}$, a target state
is observed up to a field of 0.5~T, whereas this critical field is close to the
field where target states cannot be stabilized, which is around
$1.2\,\text{T}$, when ferromagnetic boundaries are present.

A comparison of the energy of skyrmions and target states reveals that with
free boundaries target states have always larger energy than skyrmions up to
sizes of around $R=8\,\text{nm}$. In contrast, with a ferromagnetic rim target
states have lower energy than skyrmions at $R=8\,\text{nm}$ and possibly become
the ground state of the system at a critical field. This supports the hypothesis that
confinement has a larger influence on target states, most likely because of the
DMI energy and the repulsion from the boundary spins, which is weaker for a
ferromagnetic rim. Additionally, we see that at $R\geq10\,\text{nm}$ skyrmions
turn into worm or stripe domains below ${B_{z}=0.7\,\text{T}}$, however target
states are stable down to zero field since their double $360^\circ$ spin
rotation allows them to cover a larger area without distortion. 

Similar to the behaviour of isolated skyrmions in extended systems, by
increasing the magnetic field strength the skyrmion size decreases. To quantify
this change we calculated the skyrmion radius as a function of the magnetic
field for hexagon sizes between~3 and $20\,\text{nm}$ and different boundary
conditions.  This result is depicted in
Fig.~\ref{fig:sks-tgt-st-radius-vs-field}(a) and~(b). We define the skyrmion
radius as the distance from the skyrmion center to the ring where $m_{z}=0$. In
the case of $R\geq10\,\text{nm}$, the curve is not shown below certain field
magnitudes since the system relaxed into elongated domains or spiral domains.
To compare the effect of the confinement on the skyrmions with the skyrmion
size in extended samples, we additionally plot in
Fig.~\ref{fig:sks-tgt-st-radius-vs-field}(a) and~(b), as a dashed line, an
estimation of the skyrmion radius as a function of the magnetic field, based on
the continuum theory proposed in Ref.~\onlinecite{Wang2018}, which depends on
the skyrmion radius and the width of the wall separating the skyrmion core from
the ferromagnetic background.  This semi-analytical result (from equations~11
and~$12$ in Ref.~\onlinecite{Wang2018}) slightly underestimates the skyrmion
radius~\cite{Wang2018}, in part because it assumes the skyrmion radius is much
larger than the width of its enclosing wall, nevertheless it correctly
reproduces the skyrmion size dependence with $B_{z}$ and agrees reasonably well
with the skyrmion sizes shown in Ref.~\onlinecite{Romming2015}.  For hexagon
sizes below 8~nm and sufficiently small fields, we notice that skyrmions in the
system with a ferromagnetic rim (Fig.~\ref{fig:sks-tgt-st-radius-vs-field}(a))
are larger in size than the skyrmions in the sample with free boundaries
(Fig.~\ref{fig:sks-tgt-st-radius-vs-field}(b)) at equivalent field strengths
and $R$. The reason is that for sufficiently weak applied fields the skyrmion
tries to occupy the largest possible area in the island and when the island has
free boundaries the tilted spins at the hexagon edges restrict the skyrmion
size (see Supplemental Sec.~S11~\cite{suppmat}), as shown in
Fig.~\ref{fig:sks-tgt-st-radius-vs-field}(a). This is in contrast to the system
of Fig.~\ref{fig:sks-tgt-st-radius-vs-field}(b), where pinned spins at the
boundary do not limit the skyrmion and allow it to expand further towards the
sample edges. When increasing the field strength skyrmions shrink, hence for
sufficiently large sample size and applied field the boundary spins no longer
have an influence on the skyrmion and their sizes converge to similar values
regardless of the edge condition, as shown by a comparison with the theoretical
curve. Moreover, we observe that with a ferromagnetic rim the skyrmion size
behaves as in an extended film in a larger range of magnetic fields for hexagon
sizes as small as 4~nm and, in particular, for islands with $R$ around
$8\,\text{nm}$. The influence of the boundary on the skyrmion size is in
agreement with the skyrmion energies of
Fig.~\ref{fig:hexagons-R-energy-skyrmion}, thus we can say that confinement
effects on a single skyrmion are important at sufficiently weak fields for
hexagonal islands of $R\leq10\,\text{nm}$, and are stronger in islands with
free boundaries.

\begin{figure*}[p!]
    \includegraphics[width=0.9\textwidth]{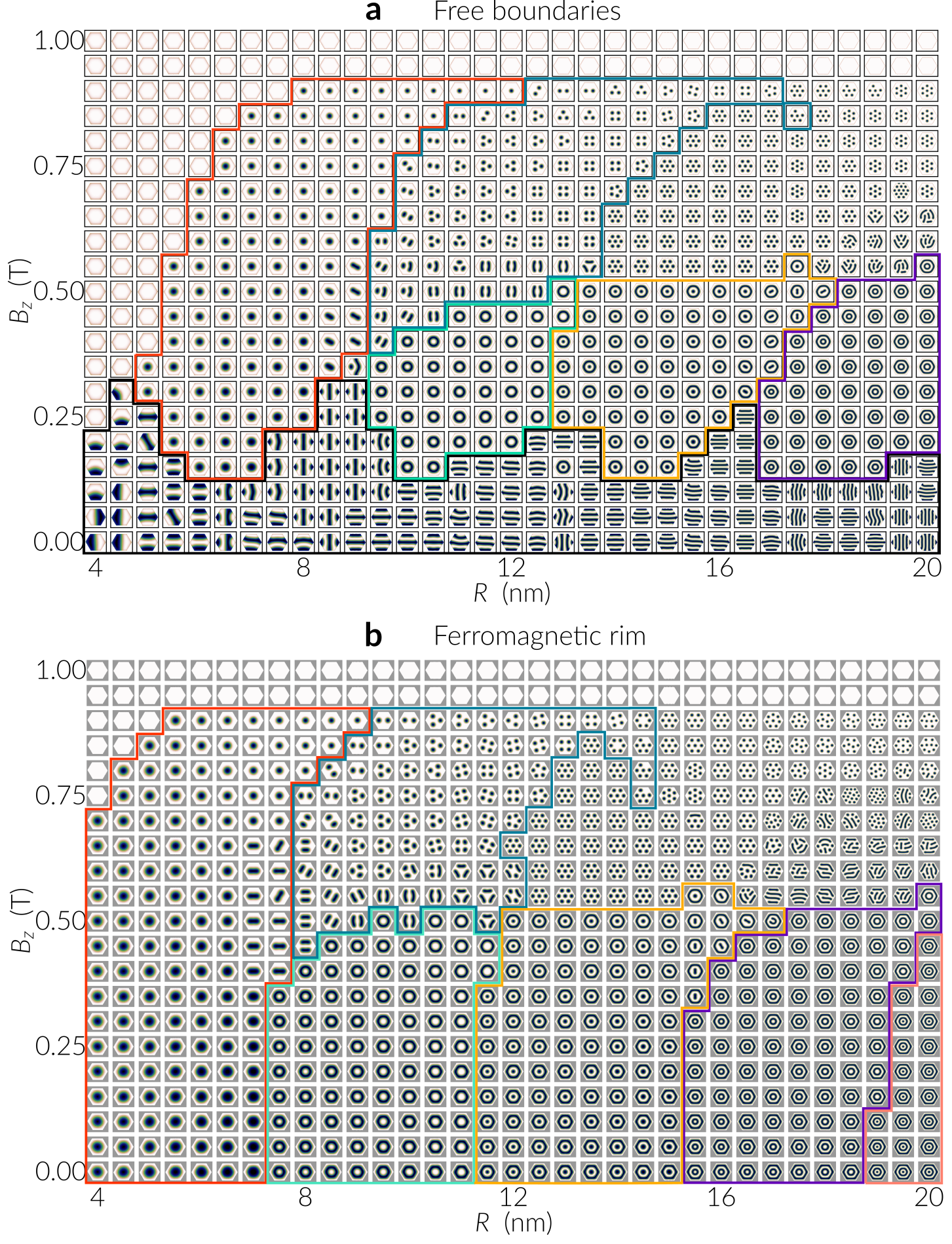}

    \caption{Lowest energy state phase diagram of the Pd/Fe/Ir(111) and Pd$_2$/Fe/Ir(111) hexagon systems. The figure shows the lowest energy states of hexagon islands as a function of their circumradius $R$ and the applied out-of-plane magnetic field $B_{z}$. Phases are marked by enclosed regions in thick lines. (a) Diagram
    of hexagons with open boundaries. (b) Diagram of hexagons with a ferromagnetic rim.}

    \label{fig:hexagons-phase-diagram}
\end{figure*}

We define the target state radius as the outermost distance where $m_{z}=0$ and
we plot it as a function of hexagon size and magnetic field in
Fig.~\ref{fig:sks-tgt-st-radius-vs-field}(c) and~(d). An analysis of the radius of
target states in hexagons with different boundary conditions results in the
same tendency exhibited by skyrmions, i.e.\ target states become smaller with
increasing field magnitudes and are larger in size when spins are not tilted at
the sample edges. However, target states are, overall, noticeable bigger than
skyrmions for hexagon radii up to $28\,\text{nm}$ and below the critical field
of 1.2~T where they become unstable.  Target states can be stabilized in
hexagons as small as $R=6\,\text{nm}$, where they are stable in a larger range
of field magnitudes when a ferromagnetic rim is present
(Fig.~\ref{fig:sks-tgt-st-radius-vs-field}(d)). Below this circumradius the
system relaxes to a skyrmion for the two boundary conditions. Furthermore, a
critical magnetic field is observed at $0.8\,\text{T}$, where the size of
target states in islands larger than $R>8\,\text{nm}$ converges to a radius
around $5\,\text{nm}$, and the configuration becomes small and well confined
within the island. By increasing the field, the radius of target states tend to
a similar value, smaller than $3\,\text{nm}$, independent of the system size.

\subsection{Phase diagram}
\label{sec:phase-diagram}

As observed in experimentally measured islands, spin spiral configurations
are the most probable ground state of the system at weak fields. This is
illustrated, in part, in Fig.~\ref{fig:island-energies-Q}, where the data point
for the state from Fig.~\ref{fig:Romming-field-sweep}(a) has the lowest energy
at zero field. Furthermore, the favored formation of spirals at weak fields
is expected from the material parameters (see Sec.~\ref{sec:exp-obs-mag-states}),
which specify a critical DMI value~\cite{Rohart2013,Mulkers2016} of $D_{\text{crit}}=4\sqrt{A
K_{\text{eff}}}/\pi=2.85\,\text{mJ m}^{-2}$, thus
$D_{\text{c}}/D_{\text{crit}}\approx 1.37$. Since the DMI magnitude is well
above the critical value, spiral solutions are energetically favored, as
can be seen in Figs.~\ref{fig:Romming-field-sweep} and~\ref{fig:hexagons-phase-diagram}(a). 

At larger field values, spin spirals are no longer favored and
skyrmionic textures are candidates for being the ground state.  In
Fig.~\ref{fig:hexagons-phase-diagram} we show full phase diagrams depicting the
lowest energy state of hexagons as a function of the circumradius $R$, the
applied field and the two different boundary conditions. The lowest energy
configurations were found by simulating 27 different initial states per $(R,B)$
point in the diagram, which correspond to multiple skyrmion states up to a
number of seven, $k\pi$-skyrmion states from $2\pi$ up to $5\pi$, random
configurations, $360^{\circ}$ domain walls and skyrmions next to a
$360^{\circ}$ wall.

The results illustrated in Fig.~\ref{fig:hexagons-phase-diagram}(a) at zero
field, for hexagons with free boundaries, confirm our observations of spin
spirals as ground states. The number of cycloids of these lowest energy
spirals, fitting within an island, depends on the island size, which is
consistent with Ref.~\onlinecite{Mulkers2016}, and their period remains
constant since, for example, islands with $R=8$ and 12~nm have twice and three
times, respectively, the number of cycloids than the case $R=4\,\text{nm}$.
Additionally, they do not seem affected by the applied field. By increasing
the field, between $R=5\,\text{nm}$ and $12\,\text{nm}$, at the top left
enclosed region we distinguish a single skyrmion phase.  The transition from
spirals into single skyrmion states is clearly dependent on the Zeeman
interaction which starts to dominate over the DMI.

Next to the single skyrmion region, at sufficiently large magnetic fields, we
marked the area where worm domains or up to~6 multiple skyrmions become the
lowest energy state. At the right of this region, a larger number of skyrmions
are the lowest energy configurations. Since we specified up to seven skyrmions
as initial states, it is likely that a slightly larger number of skyrmions have
lower energy in this region. For example, at $R=19\,\text{nm}$ and
$B_{z}=0.7\,\text{T}$ we notice a state with 12~skyrmions and~2 half-skyrmions,
whose topological charge is about -12.6. Multiple skyrmion states are favored
as the island size and applied field increase because there is enough distance
to avoid the inter-skyrmion repulsion that can overcome the energy barrier from
the boundary. The optimal arrangement of skyrmions fitting within an island
clearly follows a close packing ordering of congruent circles, as has been
discussed in Ref.~\onlinecite{Zhao2016} for the case of disks, and the system
size affects the number of confined
skyrmions,~\cite{Beg2015,Zhao2016,Pepper2018,Ho2019} which can be characterized
by the total topological charge. Furthermore, the shape of the magnetic
structures usually become distorted at the phase boundaries, in particular
skyrmions, which appear as worm domains. These phase transitions are driven by
(i) the hexagonal shape of the island, (ii) the interplay of the DMI and the
Zeeman interaction when the field changes (i.e. along the vertical
axis of Fig.~\ref{fig:hexagons-phase-diagram}), as in the case of a skyrmion in
Fig.~\ref{fig:island-energies-Q}, and (iii) by the size of the island.

Below the skyrmion regions, at lower magnetic fields and with increasing
hexagon radii, target states, $3\pi$-skyrmion and $4\pi$-skyrmion states have
the lowest energy as the DMI energy is more dominant than the Zeeman energy
and there is enough available space for these states. In this context, for
sufficiently large island sizes there is a possibility that more complex
configurations such as multiple target states and so-called skyrmion
bags~\cite{Foster2019,Rybakov2019} have lower energy. Interestingly, we
observed a few of these states during our energy minimization process and none
of them had lower energy than the states obtained from our set of initial
states. 

\begin{figure}[t!]
    \includegraphics[width=\columnwidth]{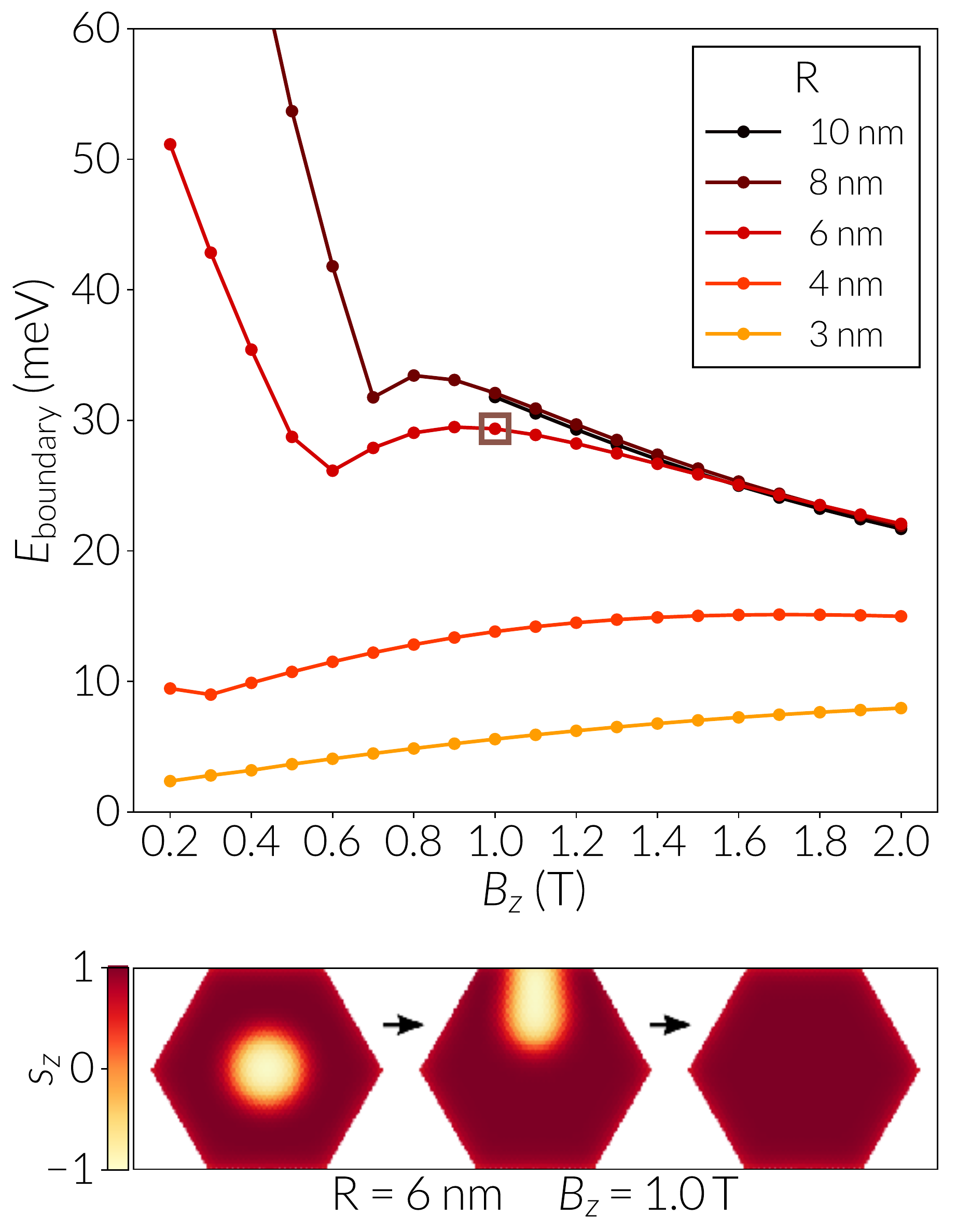}

    \caption{Energy barrier for the skyrmion escape through the boundary of
        hexagons of different size. Barriers are calculated for free boundary
        hexagons of variable circumradius $R$ and magnetic field strength. For
        $R\geq6\,\text{nm}$ and weak fields the change in the tendency of the
        curves is because the skyrmion (which has a large size at weak fields)
        elongates until reaching the boundary instead of being displaced while
        keeping its symmetrical shape. Snapshots with the escape transition are
        shown below the figure for the parameters of the data point marked with an
        open square in the energy barrier curves.}

    \label{fig:hexagons-nebm-sk-escape}
\end{figure}

\begin{figure*}[t!]
    \includegraphics[width=1.0\textwidth]{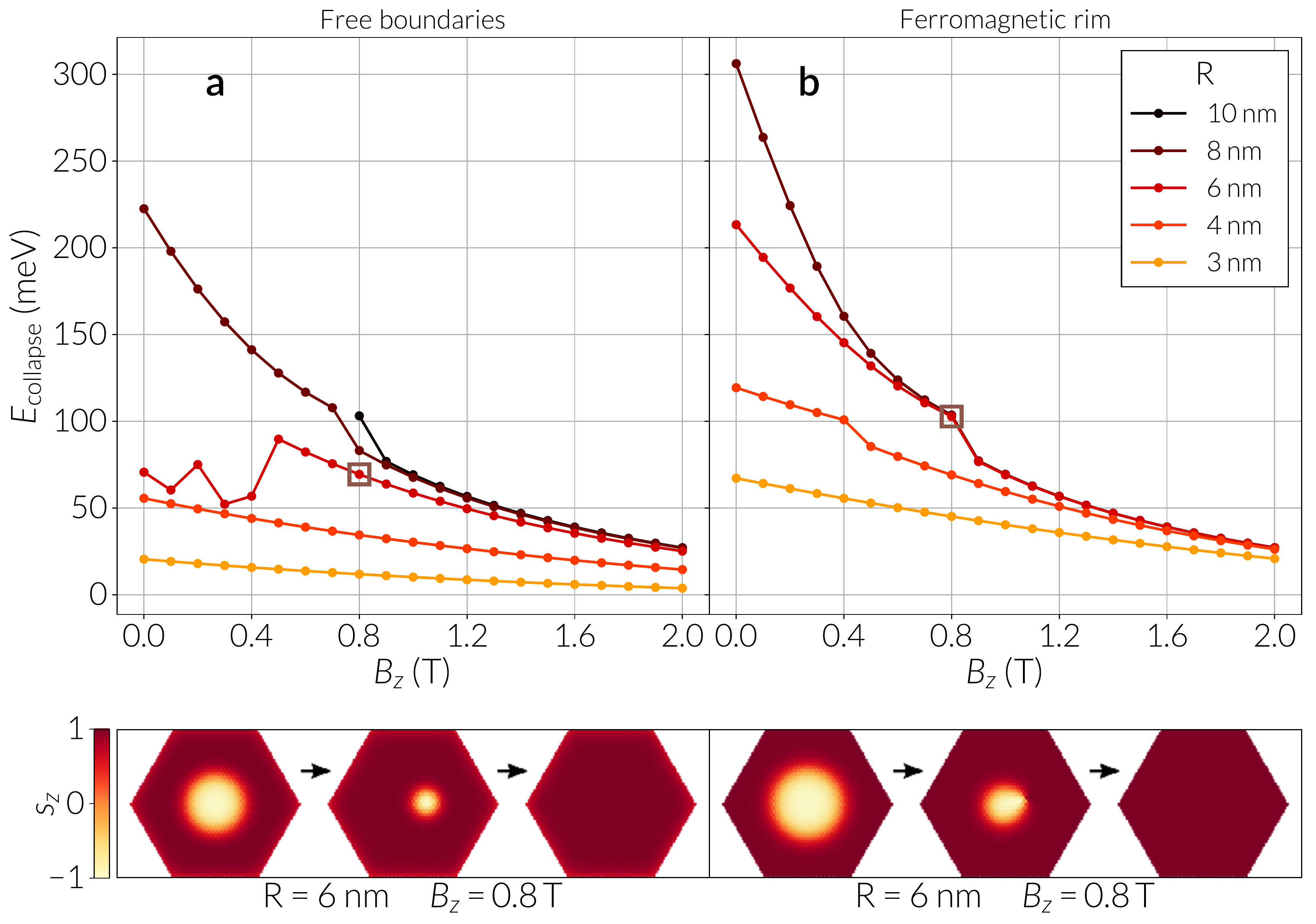}

    \caption{Energy barrier for the skyrmion collapse in hexagons of different
        size. Barriers are calculated for hexagons with either free~(a) or
        ferromagnetic~(b) boundaries, and variable circumradius $R$ and
        magnetic field strength. For $R=6\,\text{nm}$ and weak fields, the
        change in the curve tendency is because the GNEBM finds a boundary
        mediated transition rather than a collapse.  For $R\leq8\,\text{nm}$
        and weak fields, the change in the curve is because the GNEBM finds a
        skyrmion collapse mediated by a singularity. Snapshots with the
        collapse transition (below (a)) and the collapse via a singularity
        (below (b)) are shown below the corresponding plots with different
        boundary conditions and for the data points marked with open
        squares in the energy barrier curves.
    }

    \label{fig:hexagons-nebm-sk-collapse}
\end{figure*}

\begin{figure*}[t!]
    \includegraphics[width=1.0\textwidth]{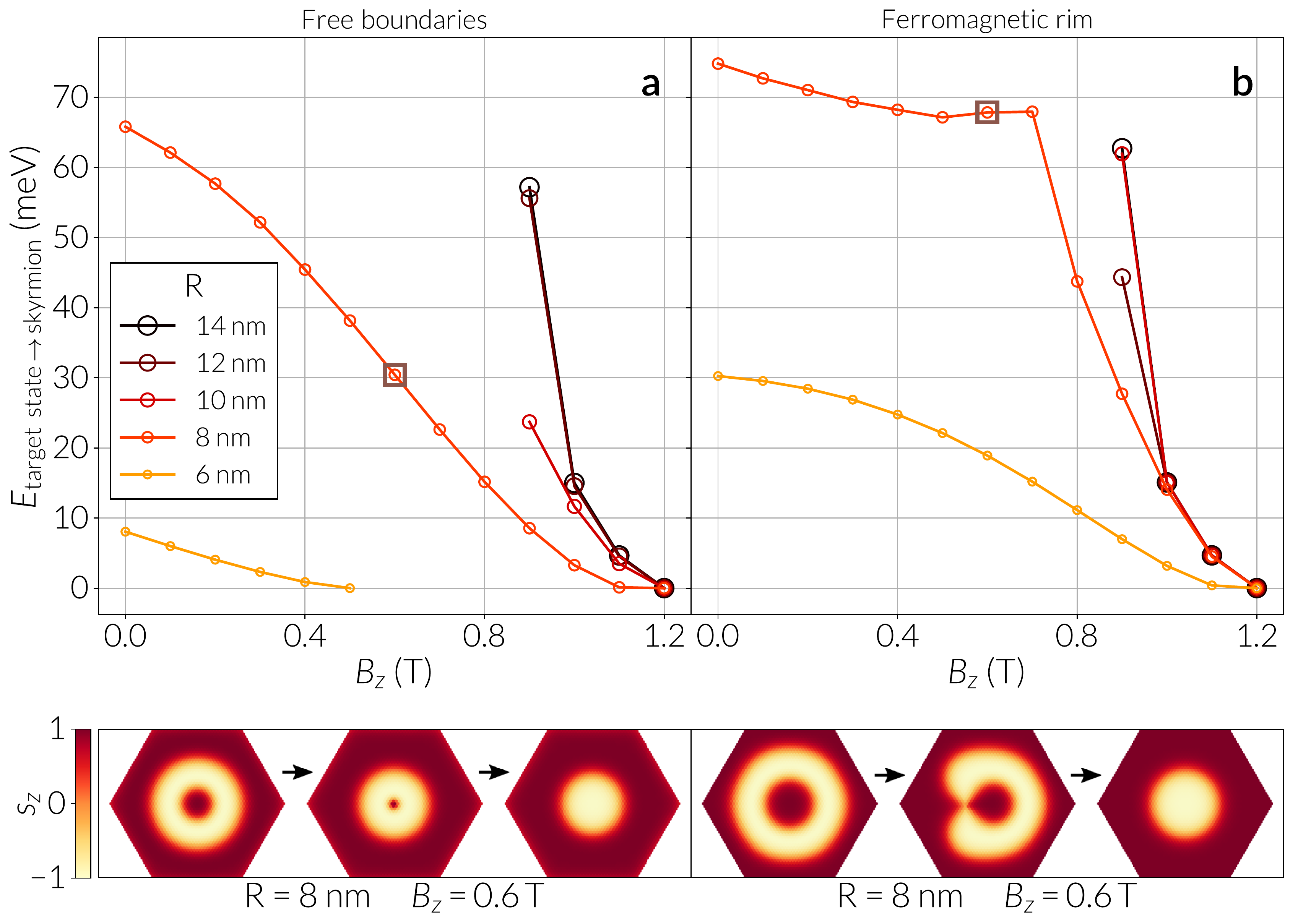}

    \caption{Energy barrier for the transition of a target state into a
        skyrmion in hexagons of different size. Barriers are calculated for
        hexagons with either free~(a) or ferromagnetic~(b) boundaries, and
        variable circumradius $R$ and magnetic field strength.  For
        sufficiently large hexagon sizes the algorithm was not able to converge
        since transitions are mediated by singularities and the initial state
        used is not optimal to find a different path.  Snapshots with an
        internal collapse of the target state (below (a)) and the transition
        via a singularity (below (b)) are shown below the corresponding plots
        with different boundary conditions, and for the data points marked
        with open squares in the energy barrier curves.
    }

    \label{fig:hexagons-nebm-tgt-state-sk}
\end{figure*}

At fields above $0.9\,\text{T}$ the ferromagnetic state is always the ground
state. At weaker fields and when the hexagon radii are smaller than 8~nm, a
region is observed where the uniform state is the lowest energy configuration
because boundary spins inhibit the formation of skyrmions.

When ferromagnetic boundaries are present spin spirals are suppressed and the
phases of skyrmionic textures significantly expand and are observed down to
zero field, as shown in Fig.~\ref{fig:hexagons-phase-diagram}(b). In particular,
a single skyrmion becomes the ground state for islands down to $4\,\text{nm}$
radius in size. Furthermore, because of the lack of tilted boundary spins,
close to $R=20\,\text{nm}$, a phase emerges with $5\pi$-skyrmion
configurations. In agreement with the case of open boundaries, the saturation
field remains 0.9~T. In this case, the region where the uniform state is the
ground state, for fields below $0.9\,\text{T}$, is smaller than in the case of
open boundaries and confirms that the tilted spins at the edge hinder the
formation of skyrmions in sufficiently small hexagons.

In Fig.~\ref{fig:hexagons-R-energy-skyrmion} we saw that for hexagons with a
radius around $8\,\text{nm}$, a ferromagnetic rim causes target states to have
smaller energy than skyrmions for fields below $0.6\,$T. From
Fig.~\ref{fig:hexagons-phase-diagram}(b) we notice that in islands with a
ferromagnetic rim the target state becomes the lowest energy state in a region
of hexagon radii from approximately 7.5~nm up to 11~nm and below fields of
0.5~T, and hence it is energetically favored in this region of phase space.
Furthermore, since the islands measured experimentally have a radius of about
9~nm we can infer that target states in Pd$_2$/Fe islands have a larger
probability of being observed if the system overcomes energy barriers towards
lowest energy configurations.

\subsection{Stability of skyrmions and target states}
\label{sec:stability-sk-hexagons}

A method for the estimation of the stability of skyrmions is the calculation of energy
barriers separating them from other equilibrium states, in particular from the uniform
state. Two known
mechanisms~\cite{Cortes2017,Stosic2017,Bessarab2015,Bessarab2018} for the
skyrmion annihilation (creation) are the skyrmion collapse (emergence) and the
skyrmion escape (nucleation) mediated by the boundary. The latter is only possible
when free boundaries are present.

In Fig.~\ref{fig:hexagons-nebm-sk-escape} we show the energy barriers between
the skyrmion and the uniform state via the escape mechanism for the system with
open boundaries. To obtain this transition, we initialized the algorithm by
moving the skyrmion towards the upper boundary of the hexagon. In the figure we
observe that for $R<6\,\text{nm}$ the height of the barrier increases with larger
field strengths, and for $R\geq6\,\text{nm}$ we see the opposite tendency.
Furthermore, for  $R\geq6\,\text{nm}$ and fields $0.6\,\text{T}$ there is a
drastic change in the slope of the curve because the skyrmion elongates
towards a boundary until reaching the sample border and starts escaping, instead
of displacing without deformation towards the sample edge, as occurs for
sufficiently large fields. In small samples the effect is unnoticed since the
skyrmion is directly touching the sample boundaries, thus it can easily
transition towards the uniform state. Snapshots of these transitions for an
$R=8\,\text{nm}$ hexagon and energy bands from the simulations are shown in Sec.~S12
of the Supplemental Material.~\cite{suppmat}

Regarding the collapse mechanism, we summarize the energy barrier
calculations in Fig.~\ref{fig:hexagons-nebm-sk-collapse}, for the two boundary
conditions. In this case we notice the barriers of hexagons with free
boundaries Fig.~\ref{fig:hexagons-nebm-sk-collapse}(a)) are smaller in magnitude
than the corresponding barriers from the system with ferromagnetic boundaries
(Fig.~\ref{fig:hexagons-nebm-sk-collapse}(b)). Additionally, in
Fig.~\ref{fig:hexagons-nebm-sk-collapse}(a) at $R=6\,\text{nm}$ and below
$B_{z}=0.5\,\text{T}$ the algorithm finds a different energy path, which is
mediated by the boundary and is not allowed when the system has fixed spins at
the boundary. The change in the curve around $0.8\,\text{T}$ for circumradius
above $6\,\text{nm}$ indicate the collapse of the skyrmion through a
singularity,~\cite{Cortes2017} which is not present at larger magnetic fields
or in small sample sizes since the singularity is defined in a few atomic
spaces. For the hexagon with the ferromagnetic rim there are no effects from
the boundary, however singularities still appear at sufficiently large samples
or skyrmion sizes, which can be seen from the small jumps in the curves of
Fig.~\ref{fig:hexagons-nebm-sk-collapse}(b). According to these results,
it is possible that the boundary has an influence on the formation of a
singularity for large enough hexagons. For fields below $0.8\,\text{T}$ and
circumradius of $10\,\text{nm}$ and above, we do not show data points since the system
relaxes into an elongated skyrmionic domain or spiral domain.

An investigation of the transition energies of target states into skyrmions can
provide us an alternative answer for the problem of stability of target states.
We computed this transition for different hexagon sizes and boundary conditions
and summarize the results in Fig.~\ref{fig:hexagons-nebm-tgt-state-sk}. From
the simulations with $R\leq8\,\text{nm}$ we observe that the minimum energy
path is given by the collapse of the inner core of the target state, which is
oriented along the field direction $+z$, making the ring of spins in the $-z$
direction to expand into the skyrmion centre (see Supplemental Fig.~S13~\cite{suppmat}). The
same mechanism has been reported by Hagemeister {\it et
al.}~\cite{Hagemeister2018} in infinite samples.  When the system has free
boundaries, other possible energy paths for the target state involve the
boundaries of the hexagon and during this process complex spiral orderings
mediate the transition. Unfortunately, our attempts to calculate transitions
from target states to either a ferromagnetic state or a skyrmion, via the
boundary of the system, were not successful owing to the lack of convergence of
the algorithm to a consistent result, thus a study of these paths requires a
more careful analysis. 

The inner collapse transitions are given by the smooth curves of
Fig.~\ref{fig:hexagons-nebm-tgt-state-sk}. An exception occurs for
$R\geq8\,\text{nm}$, $B\lessapprox1.0\,\text{T}$ and ferromagnetic boundaries
(Fig.~\ref{fig:hexagons-nebm-tgt-state-sk}(b)), since the transition in those
cases is given by the emergence of a singularity at the ring of $-z$ spins (see
Supplemental Fig.~S14~\cite{suppmat}). The reason is that the target state is larger in size
when a ferromagnetic rim is present, as we saw in
Fig.~\ref{fig:sks-tgt-st-radius-vs-field}, which must be around a critical size
for singularities to mediate the transition, as in the case of the skyrmion
collapse. The same effect occurs for $R>8\,\text{nm}$ and weak fields for the
two boundary conditions, hence the algorithm did not converge to an optimal
path for fields below $0.9\,\text{T}$.  The energy barrier calculations
indicate that target states have associated larger barriers to decay into the
skyrmion state in hexagons with a ferromagnetic rim. Moreover, the most evident
mechanism for the annihilation of the target state is through a collapse or a
disruption mediated by a singularity, since boundary effects are forbidden.
Although there are possibly multiple other energy paths for the annihilation of
a target state, for example the transition of a target state into more than one
skyrmion, they are likely to have associated larger energy barriers because
multiple singularities would be necessary to mediate these transitions. These
results suggest that the transition path of a target state into a single
skyrmion is the shortest in energy landscape and with the highest probability
to occur.

\section{Discussion}

Using (spin-polarized) scanning tunneling microscopy we have studied the
magnetic properties of two different systems, namely Pd/Fe and Pd$_2$/Fe on
Ir(111), in a confined geometry. We find that the investigated islands host a
variety of spin textures, including different metastable states. In islands
with an effectively non-magnetic surrounding the spin texture exhibits a tilt
at the edges. A strong hysteresis is observed in magnetic field dependent
measurements, where the virgin curve magnetic states are qualitatively different
from the ones obtained when the magnetic field is subsequently applied in the
opposite direction: whereas in the first case magnetic skyrmions are found, in
the latter case the magnetic states are characterized by branched $360^{\circ}$
walls which end at island boundaries. When the surrounding of an island is
ferromagnetic, such $360^{\circ}$ walls perpendicular to island edges are
avoided and target-like configurations can be found.

Using discrete spin simulations we can reproduce our experimental observations
of multiple chiral configurations in small Pd/Fe islands. An analysis of the
energy of chiral textures, which include the skyrmion, the target state and the
$3\pi$-skyrmion, indicates that these configurations can be stabilized from
zero field up to different critical field magnitudes. Moreover, from the energy
curves of these configurations as a function of the applied field, it can be
inferred that the stability of some configurations can be explained by
measuring the energy barriers separating them from energetically close
configurations, for example a target state with a skyrmion. These results can
be used as a reference for the observation of these magnetic states in future
experiments. In particular, there is still a lack of experimental evidence of
the $3\pi$-skyrmion in a small confined system with interfacial DMI.

In order to reproduce different magnetic textures from SP-STM images of a field
sweep experiment on Pd/Fe islands, we performed field sweep simulations in a
range of field magnitudes smaller than the one used in the experiment and using
specific initial states based on different numbers of skyrmions. Our simulation
results showed multiple skyrmion-like domains trapped within branched spiral
states that appear when the direction of the field is inverted after crossing
the zero field point, and we provided arguments to identify them by their
topological number. The field magnitudes for the observation of chiral
structures in our simulations differ from the ones in experiments because
simulations are performed at zero temperature without any other external
excitations such as the ones generated by the tunnel electrons in SP-STM.
Accordingly, we did not exactly reproduce the results from the experiment,
nonetheless it was still possible to stabilize most of the observed magnetic
structures from the different field stages.  Our simulations show that all the
observed magnetic configurations should be accessible through an appropriate
excitation of the system, thus further evidence of target states or
$3\pi$-skyrmions could be obtained in future experimental studies.

Using our simulations of a field sweep process it was not possible to obtain a
target state from spiral configurations but mostly skyrmions from $360^\circ$
domain walls at increasing strong field magnitudes (above $2\,\text{T}$) or
large abrupt field changes. A possible solution to obtain these axially
symmetric configurations with the field sweep method is specifying a suitable
initial state, however this might be difficult in practice.

We performed a thorough analysis of skyrmions and target states under different
conditions, by varying the size, applied field and boundary conditions of
perfectly shaped hexagons. The main motivation for the modification of the
boundary condition was the observation of a target-like state in the Pd$_2$/Fe
islands on Ir(111), where the surrounding of the islands is ferromagnetic.
Simulations of hexagonal islands of this material showed that for sufficiently
small systems, skyrmions and target states in islands with a ferromagnetic rim
have lower energy than the uniform state in a broader range of magnetic fields.
In particular, for islands with radii around $8\,\text{nm}$, target states have
lower energy than skyrmions when a ferromagnetic boundary is present.

In hexagons with free boundaries, we found that for radii up to approximately
$8\,\text{nm}$ and fields below $1.5\,\text{T}$, the tilting of spins at the
island rim decreases the DMI energy and affects the skyrmion structure, by
decreasing the skyrmion size in comparison to hexagons with a ferromagnetic
rim. 

In contrast, due to the lack of canted spins at the island rim when
ferromagnetic boundaries are present, skyrmions and target states are, in
general, larger in size compared to hexagons with open boundaries. We observe
that this larger size had an influence on the energy barriers separating the
skyrmion from the ferromagnetic ordering and the target state from the
skyrmion, making the barriers larger in systems with a ferromagnetic rim.

To gain an understanding of the lowest energy states of the hexagon systems we
computed a full phase diagram illustrating regions in hexagon radius and
applied field space with the possibly lowest energy configurations; there is no
guarantee that other complex magnetic states that could not be mapped with our
method can have lower energy. On the other hand, our random set of states
showed that other configurations beyond $k\pi$-skyrmions are usually highly
energetic, thus finding the true ground states requires a dedicated analysis
that goes beyond the scope of this work. We confirmed that spin spiral
configurations are the lowest energy states at weak fields and we identified
regions where multiple $k\pi$-skyrmion states are the lowest energy states at
sufficiently large island sizes and applied fields. From the comparison of the
energies of target states and skyrmions at hexagon sizes around $9\,\text{nm}$,
we confirm that target states become the lowest energy states when a
ferromagnetic rim is present, supporting our experimental observations of the
favored stabilization of target states in Pd$_2$/Fe islands.

Furthermore, by analyzing minimum energy paths we found that in hexagons with
free boundaries, up to fields of approximately 1.6~T and for hexagon
circumradii above $4\,\text{nm}$, the minimum energy path between skyrmions and
the uniform ordering is via the boundary, which is consistent with recent
studies.~\cite{Cortes2017,Stosic2017,Bessarab2018} A ferromagnetic rim inhibits
a skyrmion and a target state to escape through the boundary, hence forcing
them to transition by a collapse or by a singularity.  At weak fields and
sufficiently large sizes these transitions have an energy barrier significantly
larger than an escape mediated by the boundary when spins are not pinned at the
island edges.  Moreover, early results have shown that with free boundaries a
target state can also transition at the boundary through spiral configurations.
For instance, a lollipop-like state is formed when a target state reaches the
sample edge and a single $360^{\circ}$ wall emerges when half of the target
state has escaped. These configurations are energetically favored at weak
magnetic fields and as we saw in the simulated field sweep process, they tend
to stabilize with branches perpendicular to the sample edges.  These arguments
indicate that skyrmions and target states should be more stable in islands with
ferromagnetic boundaries since there is a larger energy cost for these
structures to transition into other accessible states in the energy landscape.
In this context, a more detailed insight into the lifetime of the observed
configurations would require the calculation of transition rates, as in
Ref.~\onlinecite{Malottki2019}. Furthermore, an accurate model of these solutions
could be obtained in the future using effective magnitudes for the material
parameters of Pd$_2$/Fe/Ir(111). For example, recently
Refs.~\onlinecite{Dupe2014,Bouhassoune2019} have reported magnitudes from {\it
ab initio} calculations.

\section{Conclusions}

In summary, in this work we have performed SP-STM measurements of skyrmionic
textures in Pd/Fe and Pd$_2$/Fe islands on Ir(111) and we have employed
discrete spin simulations to analyze our experimental observations. Our
results show that discrete spin simulations can accurately explain our
experimental findings. Furthermore, by modelling hexagonally shaped islands we
provide a more general understanding of how the applied field, system size and
boundary conditions affect the stability of skyrmions and target states in
confined geometries with interfacial DMI. In particular, we found that the
presence of ferromagnetic boundaries favor the stability of target states in a
region of applied field and island size that explains the experimental
observation of a target-like state in Pd$_{2}$/Fe/Ir(111) islands. In
addition, simulations predict conditions for the stability of novel magnetic
configurations, such as $k\pi$-skyrmions, that could be experimentally observed
in the future. 

We believe that the ease to carry out multiple computer simulated experiments
can help for the future design of energy efficient spintronic devices based on
small confined DMI systems, such as bit patterned recording media, where
skyrmions or target states would encode the memory bits.


\section{Acknowledgments}

We acknowledge financial support from {CONICYT} Chilean scholarship programme
Becas Chile (72140061), the EPSRC Complex Systems Simulations DTC
(EP/G03690X/1), the EPSRC Programme grant on {Skyrmionics} (EP/N032128/1), the
Horizon 2020 European Research Infrastructure project OpenDreamKit (67 6541),
and by the {Deutsche} {Forschungsgemeinschaft} (DFG, German Research Foundation) -
402843438; SFB668-A8.

The simulation data for a complete reproduction of this work can be found in
Ref.~\onlinecite{Cortes2019}.


\bibliography{minimal_lib}

\begin{thebibliography}{55}%
\makeatletter
\providecommand \@ifxundefined [1]{%
 \@ifx{#1\undefined}
}%
\providecommand \@ifnum [1]{%
 \ifnum #1\expandafter \@firstoftwo
 \else \expandafter \@secondoftwo
 \fi
}%
\providecommand \@ifx [1]{%
 \ifx #1\expandafter \@firstoftwo
 \else \expandafter \@secondoftwo
 \fi
}%
\providecommand \natexlab [1]{#1}%
\providecommand \enquote  [1]{``#1''}%
\providecommand \bibnamefont  [1]{#1}%
\providecommand \bibfnamefont [1]{#1}%
\providecommand \citenamefont [1]{#1}%
\providecommand \href@noop [0]{\@secondoftwo}%
\providecommand \href [0]{\begingroup \@sanitize@url \@href}%
\providecommand \@href[1]{\@@startlink{#1}\@@href}%
\providecommand \@@href[1]{\endgroup#1\@@endlink}%
\providecommand \@sanitize@url [0]{\catcode `\\12\catcode `\$12\catcode
  `\&12\catcode `\#12\catcode `\^12\catcode `\_12\catcode `\%12\relax}%
\providecommand \@@startlink[1]{}%
\providecommand \@@endlink[0]{}%
\providecommand \url  [0]{\begingroup\@sanitize@url \@url }%
\providecommand \@url [1]{\endgroup\@href {#1}{\urlprefix }}%
\providecommand \urlprefix  [0]{URL }%
\providecommand \Eprint [0]{\href }%
\providecommand \doibase [0]{http://dx.doi.org/}%
\providecommand \selectlanguage [0]{\@gobble}%
\providecommand \bibinfo  [0]{\@secondoftwo}%
\providecommand \bibfield  [0]{\@secondoftwo}%
\providecommand \translation [1]{[#1]}%
\providecommand \BibitemOpen [0]{}%
\providecommand \bibitemStop [0]{}%
\providecommand \bibitemNoStop [0]{.\EOS\space}%
\providecommand \EOS [0]{\spacefactor3000\relax}%
\providecommand \BibitemShut  [1]{\csname bibitem#1\endcsname}%
\let\auto@bib@innerbib\@empty
\bibitem [{\citenamefont {Bogdanov}\ \emph {et~al.}(1989)\citenamefont
  {Bogdanov}, \citenamefont {Kudinov},\ and\ \citenamefont
  {Yablonskii}}]{Bogdanov1989a}%
  \BibitemOpen
  \bibfield  {author} {\bibinfo {author} {\bibfnamefont {A.~N.}\ \bibnamefont
  {Bogdanov}}, \bibinfo {author} {\bibfnamefont {M.~V.}\ \bibnamefont
  {Kudinov}}, \ and\ \bibinfo {author} {\bibfnamefont {D.~A.}\ \bibnamefont
  {Yablonskii}},\ }\href@noop {} {\bibfield  {journal} {\bibinfo  {journal}
  {Sov. Phys. Solid State}\ }\textbf {\bibinfo {volume} {31}},\ \bibinfo
  {pages} {1707} (\bibinfo {year} {1989})}\BibitemShut {NoStop}%
\bibitem [{\citenamefont {Bogdanov}\ and\ \citenamefont
  {Yablonskii}(1989)}]{Bogdanov1989}%
  \BibitemOpen
  \bibfield  {author} {\bibinfo {author} {\bibfnamefont {A.}~\bibnamefont
  {Bogdanov}}\ and\ \bibinfo {author} {\bibfnamefont {D.}~\bibnamefont
  {Yablonskii}},\ }\href
  {http://www.jetp.ac.ru/cgi-bin/e/index/e/68/1/p101?a=list} {\bibfield
  {journal} {\bibinfo  {journal} {Zh. Eksp. Teor. Fiz}\ }\textbf {\bibinfo
  {volume} {95}},\ \bibinfo {pages} {178} (\bibinfo {year} {1989})}\BibitemShut
  {NoStop}%
\bibitem [{\citenamefont {R{\"o}{\ss}ler}\ \emph {et~al.}(2006)\citenamefont
  {R{\"o}{\ss}ler}, \citenamefont {Bogdanov},\ and\ \citenamefont
  {Pfleiderer}}]{Rossler2006}%
  \BibitemOpen
  \bibfield  {author} {\bibinfo {author} {\bibfnamefont {U.~K.}\ \bibnamefont
  {R{\"o}{\ss}ler}}, \bibinfo {author} {\bibfnamefont {A.~N.}\ \bibnamefont
  {Bogdanov}}, \ and\ \bibinfo {author} {\bibfnamefont {C.}~\bibnamefont
  {Pfleiderer}},\ }\href {\doibase 10.1038/nature05056} {\bibfield  {journal}
  {\bibinfo  {journal} {Nature}\ }\textbf {\bibinfo {volume} {442}},\ \bibinfo
  {pages} {797} (\bibinfo {year} {2006})}\BibitemShut {NoStop}%
\bibitem [{\citenamefont {M{\"{u}}hlbauer}\ \emph {et~al.}(2009)\citenamefont
  {M{\"{u}}hlbauer}, \citenamefont {Binz}, \citenamefont {Jonietz},
  \citenamefont {Pfleiderer}, \citenamefont {Rosch}, \citenamefont {Neubauer},
  \citenamefont {Georgii},\ and\ \citenamefont {B{\"{o}}ni}}]{Muehlbauer2009}%
  \BibitemOpen
  \bibfield  {author} {\bibinfo {author} {\bibfnamefont {S.}~\bibnamefont
  {M{\"{u}}hlbauer}}, \bibinfo {author} {\bibfnamefont {B.}~\bibnamefont
  {Binz}}, \bibinfo {author} {\bibfnamefont {F.}~\bibnamefont {Jonietz}},
  \bibinfo {author} {\bibfnamefont {C.}~\bibnamefont {Pfleiderer}}, \bibinfo
  {author} {\bibfnamefont {A.}~\bibnamefont {Rosch}}, \bibinfo {author}
  {\bibfnamefont {A.}~\bibnamefont {Neubauer}}, \bibinfo {author}
  {\bibfnamefont {R.}~\bibnamefont {Georgii}}, \ and\ \bibinfo {author}
  {\bibfnamefont {P.}~\bibnamefont {B{\"{o}}ni}},\ }\href@noop {} {\bibfield
  {journal} {\bibinfo  {journal} {Science}\ }\textbf {\bibinfo {volume}
  {323}},\ \bibinfo {pages} {915} (\bibinfo {year} {2009})}\BibitemShut
  {NoStop}%
\bibitem [{\citenamefont {Heinze}\ \emph {et~al.}(2011)\citenamefont {Heinze},
  \citenamefont {von Bergmann}, \citenamefont {Menzel}, \citenamefont {Brede},
  \citenamefont {Kubetzka}, \citenamefont {Wiesendanger},\ and\ \citenamefont
  {Bl{\"{u}}gel}}]{Heinze2011}%
  \BibitemOpen
  \bibfield  {author} {\bibinfo {author} {\bibfnamefont {S.}~\bibnamefont
  {Heinze}}, \bibinfo {author} {\bibfnamefont {K.}~\bibnamefont {von
  Bergmann}}, \bibinfo {author} {\bibfnamefont {M.}~\bibnamefont {Menzel}},
  \bibinfo {author} {\bibfnamefont {J.}~\bibnamefont {Brede}}, \bibinfo
  {author} {\bibfnamefont {A.}~\bibnamefont {Kubetzka}}, \bibinfo {author}
  {\bibfnamefont {R.}~\bibnamefont {Wiesendanger}}, \ and\ \bibinfo {author}
  {\bibfnamefont {G.~B.~S.}\ \bibnamefont {Bl{\"{u}}gel}},\ }\href@noop {}
  {\bibfield  {journal} {\bibinfo  {journal} {Nature Physics}\ }\textbf
  {\bibinfo {volume} {7}},\ \bibinfo {pages} {713} (\bibinfo {year}
  {2011})}\BibitemShut {NoStop}%
\bibitem [{\citenamefont {Romming}\ \emph {et~al.}(2013)\citenamefont
  {Romming}, \citenamefont {Hanneken}, \citenamefont {Menzel}, \citenamefont
  {Bickel}, \citenamefont {Wolter}, \citenamefont {von Bergmann}, \citenamefont
  {Kubetzka},\ and\ \citenamefont {Wiesendanger}}]{Romming2013}%
  \BibitemOpen
  \bibfield  {author} {\bibinfo {author} {\bibfnamefont {N.}~\bibnamefont
  {Romming}}, \bibinfo {author} {\bibfnamefont {C.}~\bibnamefont {Hanneken}},
  \bibinfo {author} {\bibfnamefont {M.}~\bibnamefont {Menzel}}, \bibinfo
  {author} {\bibfnamefont {J.~E.}\ \bibnamefont {Bickel}}, \bibinfo {author}
  {\bibfnamefont {B.}~\bibnamefont {Wolter}}, \bibinfo {author} {\bibfnamefont
  {K.}~\bibnamefont {von Bergmann}}, \bibinfo {author} {\bibfnamefont
  {A.}~\bibnamefont {Kubetzka}}, \ and\ \bibinfo {author} {\bibfnamefont
  {R.}~\bibnamefont {Wiesendanger}},\ }\href {\doibase 10.1126/science.1240573}
  {\bibfield  {journal} {\bibinfo  {journal} {Science}\ }\textbf {\bibinfo
  {volume} {341}},\ \bibinfo {pages} {636} (\bibinfo {year}
  {2013})}\BibitemShut {NoStop}%
\bibitem [{\citenamefont {Romming}\ \emph {et~al.}(2015)\citenamefont
  {Romming}, \citenamefont {Kubetzka}, \citenamefont {Hanneken}, \citenamefont
  {von Bergmann},\ and\ \citenamefont {Wiesendanger}}]{Romming2015}%
  \BibitemOpen
  \bibfield  {author} {\bibinfo {author} {\bibfnamefont {N.}~\bibnamefont
  {Romming}}, \bibinfo {author} {\bibfnamefont {A.}~\bibnamefont {Kubetzka}},
  \bibinfo {author} {\bibfnamefont {C.}~\bibnamefont {Hanneken}}, \bibinfo
  {author} {\bibfnamefont {K.}~\bibnamefont {von Bergmann}}, \ and\ \bibinfo
  {author} {\bibfnamefont {R.}~\bibnamefont {Wiesendanger}},\ }\href {\doibase
  10.1103/PhysRevLett.114.177203} {\bibfield  {journal} {\bibinfo  {journal}
  {Phys. Rev. Lett.}\ }\textbf {\bibinfo {volume} {114}},\ \bibinfo {pages}
  {177203} (\bibinfo {year} {2015})}\BibitemShut {NoStop}%
\bibitem [{\citenamefont {Wiesendanger}(2016)}]{Wiesendanger2016}%
  \BibitemOpen
  \bibfield  {author} {\bibinfo {author} {\bibfnamefont {R.}~\bibnamefont
  {Wiesendanger}},\ }\href {\doibase 10.1038/natrevmats.2016.44} {\bibfield
  {journal} {\bibinfo  {journal} {Nature Reviews Materials}\ }\textbf {\bibinfo
  {volume} {1}},\ \bibinfo {pages} {16044} (\bibinfo {year}
  {2016})}\BibitemShut {NoStop}%
\bibitem [{\citenamefont {Bogdanov}\ and\ \citenamefont
  {Hubert}(1994{\natexlab{a}})}]{Bogdanov1994}%
  \BibitemOpen
  \bibfield  {author} {\bibinfo {author} {\bibfnamefont {A.}~\bibnamefont
  {Bogdanov}}\ and\ \bibinfo {author} {\bibfnamefont {A.}~\bibnamefont
  {Hubert}},\ }\href {http://www.sciencedirect.com/science/article/pii/
  0304885394900469} {\bibfield  {journal} {\bibinfo  {journal} {Journal of
  magnetism and magnetic materials}\ }\textbf {\bibinfo {volume} {138}},\
  \bibinfo {pages} {255} (\bibinfo {year} {1994}{\natexlab{a}})}\BibitemShut
  {NoStop}%
\bibitem [{\citenamefont {Bogdanov}\ and\ \citenamefont
  {Hubert}(1999)}]{Bogdanov1999}%
  \BibitemOpen
  \bibfield  {author} {\bibinfo {author} {\bibfnamefont {A.}~\bibnamefont
  {Bogdanov}}\ and\ \bibinfo {author} {\bibfnamefont {A.}~\bibnamefont
  {Hubert}},\ }\href {\doibase 10.1016/S0304-8853(98)01038-5} {\bibfield
  {journal} {\bibinfo  {journal} {Journal of Magnetism and Magnetic Materials}\
  }\textbf {\bibinfo {volume} {195}},\ \bibinfo {pages} {182} (\bibinfo {year}
  {1999})}\BibitemShut {NoStop}%
\bibitem [{\citenamefont {Rohart}\ and\ \citenamefont
  {Thiaville}(2013)}]{Rohart2013}%
  \BibitemOpen
  \bibfield  {author} {\bibinfo {author} {\bibfnamefont {S.}~\bibnamefont
  {Rohart}}\ and\ \bibinfo {author} {\bibfnamefont {A.}~\bibnamefont
  {Thiaville}},\ }\href {\doibase 10.1103/PhysRevB.88.184422} {\bibfield
  {journal} {\bibinfo  {journal} {Physical Review B}\ }\textbf {\bibinfo
  {volume} {88}},\ \bibinfo {pages} {184422} (\bibinfo {year}
  {2013})}\BibitemShut {NoStop}%
\bibitem [{\citenamefont {Leonov}\ \emph {et~al.}(2014)\citenamefont {Leonov},
  \citenamefont {R{\"o}{\ss}ler},\ and\ \citenamefont
  {Mostovoy}}]{Leonov2014a}%
  \BibitemOpen
  \bibfield  {author} {\bibinfo {author} {\bibfnamefont {A.~O.}\ \bibnamefont
  {Leonov}}, \bibinfo {author} {\bibfnamefont {U.~K.}\ \bibnamefont
  {R{\"o}{\ss}ler}}, \ and\ \bibinfo {author} {\bibfnamefont {M.}~\bibnamefont
  {Mostovoy}},\ }in\ \href@noop {} {\emph {\bibinfo {booktitle} {EPJ Web of
  Conferences}}},\ Vol.~\bibinfo {volume} {75}\ (\bibinfo {organization} {EDP
  Sciences},\ \bibinfo {year} {2014})\ p.\ \bibinfo {pages} {05002}\BibitemShut
  {NoStop}%
\bibitem [{\citenamefont {Beg}\ \emph {et~al.}(2015)\citenamefont {Beg},
  \citenamefont {Carey}, \citenamefont {Wang}, \citenamefont
  {Cort{\'{e}}s-Ortu{\~{n}}o}, \citenamefont {Vousden}, \citenamefont
  {Bisotti}, \citenamefont {Albert}, \citenamefont {Chernyshenko},
  \citenamefont {Hovorka}, \citenamefont {Stamps},\ and\ \citenamefont
  {Fangohr}}]{Beg2015}%
  \BibitemOpen
  \bibfield  {author} {\bibinfo {author} {\bibfnamefont {M.}~\bibnamefont
  {Beg}}, \bibinfo {author} {\bibfnamefont {R.}~\bibnamefont {Carey}}, \bibinfo
  {author} {\bibfnamefont {W.}~\bibnamefont {Wang}}, \bibinfo {author}
  {\bibfnamefont {D.}~\bibnamefont {Cort{\'{e}}s-Ortu{\~{n}}o}}, \bibinfo
  {author} {\bibfnamefont {M.}~\bibnamefont {Vousden}}, \bibinfo {author}
  {\bibfnamefont {M.-A.}\ \bibnamefont {Bisotti}}, \bibinfo {author}
  {\bibfnamefont {M.}~\bibnamefont {Albert}}, \bibinfo {author} {\bibfnamefont
  {D.}~\bibnamefont {Chernyshenko}}, \bibinfo {author} {\bibfnamefont
  {O.}~\bibnamefont {Hovorka}}, \bibinfo {author} {\bibfnamefont {R.~L.}\
  \bibnamefont {Stamps}}, \ and\ \bibinfo {author} {\bibfnamefont
  {H.}~\bibnamefont {Fangohr}},\ }\href {\doibase 10.1038/srep17137} {\bibfield
   {journal} {\bibinfo  {journal} {Scientific Reports}\ }\textbf {\bibinfo
  {volume} {5}},\ \bibinfo {pages} {17137} (\bibinfo {year}
  {2015})}\BibitemShut {NoStop}%
\bibitem [{\citenamefont {Liu}\ \emph {et~al.}(2015)\citenamefont {Liu},
  \citenamefont {Du}, \citenamefont {Jia},\ and\ \citenamefont
  {Du}}]{Liu2015a}%
  \BibitemOpen
  \bibfield  {author} {\bibinfo {author} {\bibfnamefont {Y.}~\bibnamefont
  {Liu}}, \bibinfo {author} {\bibfnamefont {H.}~\bibnamefont {Du}}, \bibinfo
  {author} {\bibfnamefont {M.}~\bibnamefont {Jia}}, \ and\ \bibinfo {author}
  {\bibfnamefont {A.}~\bibnamefont {Du}},\ }\href@noop {} {\bibfield  {journal}
  {\bibinfo  {journal} {Physical Review B}\ }\textbf {\bibinfo {volume} {91}},\
  \bibinfo {pages} {094425} (\bibinfo {year} {2015})}\BibitemShut {NoStop}%
\bibitem [{\citenamefont {Carey}\ \emph {et~al.}(2016)\citenamefont {Carey},
  \citenamefont {Beg}, \citenamefont {Albert}, \citenamefont {Bisotti},
  \citenamefont {Cort{\'e}s-Ortu{\~n}o}, \citenamefont {Vousden}, \citenamefont
  {Wang}, \citenamefont {Hovorka},\ and\ \citenamefont {Fangohr}}]{Carey2016}%
  \BibitemOpen
  \bibfield  {author} {\bibinfo {author} {\bibfnamefont {R.}~\bibnamefont
  {Carey}}, \bibinfo {author} {\bibfnamefont {M.}~\bibnamefont {Beg}}, \bibinfo
  {author} {\bibfnamefont {M.}~\bibnamefont {Albert}}, \bibinfo {author}
  {\bibfnamefont {M.-A.}\ \bibnamefont {Bisotti}}, \bibinfo {author}
  {\bibfnamefont {D.}~\bibnamefont {Cort{\'e}s-Ortu{\~n}o}}, \bibinfo {author}
  {\bibfnamefont {M.}~\bibnamefont {Vousden}}, \bibinfo {author} {\bibfnamefont
  {W.}~\bibnamefont {Wang}}, \bibinfo {author} {\bibfnamefont {O.}~\bibnamefont
  {Hovorka}}, \ and\ \bibinfo {author} {\bibfnamefont {H.}~\bibnamefont
  {Fangohr}},\ }\href {\doibase 10.1063/1.4962726} {\bibfield  {journal}
  {\bibinfo  {journal} {Applied Physics Letters}\ }\textbf {\bibinfo {volume}
  {109}},\ \bibinfo {pages} {122401} (\bibinfo {year} {2016})}\BibitemShut
  {NoStop}%
\bibitem [{\citenamefont {Pepper}\ \emph {et~al.}(2018)\citenamefont {Pepper},
  \citenamefont {Beg}, \citenamefont {Cort{\'e}s-Ortu{\~n}o}, \citenamefont
  {Kluyver}, \citenamefont {Bisotti}, \citenamefont {Carey}, \citenamefont
  {Vousden}, \citenamefont {Albert}, \citenamefont {Wang}, \citenamefont
  {Hovorka},\ and\ \citenamefont {Fangohr}}]{Pepper2018}%
  \BibitemOpen
  \bibfield  {author} {\bibinfo {author} {\bibfnamefont {R.~A.}\ \bibnamefont
  {Pepper}}, \bibinfo {author} {\bibfnamefont {M.}~\bibnamefont {Beg}},
  \bibinfo {author} {\bibfnamefont {D.}~\bibnamefont {Cort{\'e}s-Ortu{\~n}o}},
  \bibinfo {author} {\bibfnamefont {T.}~\bibnamefont {Kluyver}}, \bibinfo
  {author} {\bibfnamefont {M.-A.}\ \bibnamefont {Bisotti}}, \bibinfo {author}
  {\bibfnamefont {R.}~\bibnamefont {Carey}}, \bibinfo {author} {\bibfnamefont
  {M.}~\bibnamefont {Vousden}}, \bibinfo {author} {\bibfnamefont
  {M.}~\bibnamefont {Albert}}, \bibinfo {author} {\bibfnamefont
  {W.}~\bibnamefont {Wang}}, \bibinfo {author} {\bibfnamefont {O.}~\bibnamefont
  {Hovorka}}, \ and\ \bibinfo {author} {\bibfnamefont {H.}~\bibnamefont
  {Fangohr}},\ }\href {\doibase 10.1063/1.5022567} {\bibfield  {journal}
  {\bibinfo  {journal} {Journal of Applied Physics}\ }\textbf {\bibinfo
  {volume} {123}},\ \bibinfo {pages} {093903} (\bibinfo {year}
  {2018})}\BibitemShut {NoStop}%
\bibitem [{\citenamefont {Kolesnikov}\ \emph {et~al.}(2018)\citenamefont
  {Kolesnikov}, \citenamefont {Stebliy}, \citenamefont {Samardak},\ and\
  \citenamefont {Ognev}}]{Kolesnikov2018}%
  \BibitemOpen
  \bibfield  {author} {\bibinfo {author} {\bibfnamefont {A.~G.}\ \bibnamefont
  {Kolesnikov}}, \bibinfo {author} {\bibfnamefont {M.~E.}\ \bibnamefont
  {Stebliy}}, \bibinfo {author} {\bibfnamefont {A.~S.}\ \bibnamefont
  {Samardak}}, \ and\ \bibinfo {author} {\bibfnamefont {A.~V.}\ \bibnamefont
  {Ognev}},\ }\href {\doibase 10.1038/s41598-018-34934-2} {\bibfield  {journal}
  {\bibinfo  {journal} {Scientific Reports}\ }\textbf {\bibinfo {volume} {8}},\
  \bibinfo {pages} {16966} (\bibinfo {year} {2018})}\BibitemShut {NoStop}%
\bibitem [{\citenamefont {Hagemeister}\ \emph {et~al.}(2018)\citenamefont
  {Hagemeister}, \citenamefont {Siemens}, \citenamefont {R\'ozsa},
  \citenamefont {Vedmedenko},\ and\ \citenamefont
  {Wiesendanger}}]{Hagemeister2018}%
  \BibitemOpen
  \bibfield  {author} {\bibinfo {author} {\bibfnamefont {J.}~\bibnamefont
  {Hagemeister}}, \bibinfo {author} {\bibfnamefont {A.}~\bibnamefont
  {Siemens}}, \bibinfo {author} {\bibfnamefont {L.}~\bibnamefont {R\'ozsa}},
  \bibinfo {author} {\bibfnamefont {E.~Y.}\ \bibnamefont {Vedmedenko}}, \ and\
  \bibinfo {author} {\bibfnamefont {R.}~\bibnamefont {Wiesendanger}},\ }\href
  {\doibase 10.1103/PhysRevB.97.174436} {\bibfield  {journal} {\bibinfo
  {journal} {Physical Review B}\ }\textbf {\bibinfo {volume} {97}},\ \bibinfo
  {pages} {174436} (\bibinfo {year} {2018})}\BibitemShut {NoStop}%
\bibitem [{\citenamefont {Shen}\ \emph {et~al.}(2018)\citenamefont {Shen},
  \citenamefont {Zhang}, \citenamefont {Ou-Yang}, \citenamefont {Yang},\ and\
  \citenamefont {You}}]{Shen2018}%
  \BibitemOpen
  \bibfield  {author} {\bibinfo {author} {\bibfnamefont {M.}~\bibnamefont
  {Shen}}, \bibinfo {author} {\bibfnamefont {Y.}~\bibnamefont {Zhang}},
  \bibinfo {author} {\bibfnamefont {J.}~\bibnamefont {Ou-Yang}}, \bibinfo
  {author} {\bibfnamefont {X.}~\bibnamefont {Yang}}, \ and\ \bibinfo {author}
  {\bibfnamefont {L.}~\bibnamefont {You}},\ }\href {\doibase 10.1063/1.5010605}
  {\bibfield  {journal} {\bibinfo  {journal} {Applied Physics Letters}\
  }\textbf {\bibinfo {volume} {112}},\ \bibinfo {pages} {062403} (\bibinfo
  {year} {2018})}\BibitemShut {NoStop}%
\bibitem [{\citenamefont {Li}\ \emph {et~al.}(2018)\citenamefont {Li},
  \citenamefont {Xia}, \citenamefont {Zhang}, \citenamefont {Ezawa},
  \citenamefont {Kang}, \citenamefont {Liu}, \citenamefont {Zhou},\ and\
  \citenamefont {Zhao}}]{Li2018}%
  \BibitemOpen
  \bibfield  {author} {\bibinfo {author} {\bibfnamefont {S.}~\bibnamefont
  {Li}}, \bibinfo {author} {\bibfnamefont {J.}~\bibnamefont {Xia}}, \bibinfo
  {author} {\bibfnamefont {X.}~\bibnamefont {Zhang}}, \bibinfo {author}
  {\bibfnamefont {M.}~\bibnamefont {Ezawa}}, \bibinfo {author} {\bibfnamefont
  {W.}~\bibnamefont {Kang}}, \bibinfo {author} {\bibfnamefont {X.}~\bibnamefont
  {Liu}}, \bibinfo {author} {\bibfnamefont {Y.}~\bibnamefont {Zhou}}, \ and\
  \bibinfo {author} {\bibfnamefont {W.}~\bibnamefont {Zhao}},\ }\href {\doibase
  10.1063/1.5026632} {\bibfield  {journal} {\bibinfo  {journal} {Applied
  Physics Letters}\ }\textbf {\bibinfo {volume} {112}},\ \bibinfo {pages}
  {142404} (\bibinfo {year} {2018})}\BibitemShut {NoStop}%
\bibitem [{\citenamefont {Komineas}\ and\ \citenamefont
  {Papanicolaou}(2015{\natexlab{a}})}]{Komineas2015a}%
  \BibitemOpen
  \bibfield  {author} {\bibinfo {author} {\bibfnamefont {S.}~\bibnamefont
  {Komineas}}\ and\ \bibinfo {author} {\bibfnamefont {N.}~\bibnamefont
  {Papanicolaou}},\ }\href {\doibase 10.1103/PhysRevB.92.064412} {\bibfield
  {journal} {\bibinfo  {journal} {Physical Review B}\ }\textbf {\bibinfo
  {volume} {92}},\ \bibinfo {pages} {064412} (\bibinfo {year}
  {2015}{\natexlab{a}})}\BibitemShut {NoStop}%
\bibitem [{\citenamefont {Komineas}\ and\ \citenamefont
  {Papanicolaou}(2015{\natexlab{b}})}]{Komineas2015}%
  \BibitemOpen
  \bibfield  {author} {\bibinfo {author} {\bibfnamefont {S.}~\bibnamefont
  {Komineas}}\ and\ \bibinfo {author} {\bibfnamefont {N.}~\bibnamefont
  {Papanicolaou}},\ }\href@noop {} {\bibfield  {journal} {\bibinfo  {journal}
  {Physical Review B}\ }\textbf {\bibinfo {volume} {92}},\ \bibinfo {pages}
  {174405} (\bibinfo {year} {2015}{\natexlab{b}})}\BibitemShut {NoStop}%
\bibitem [{\citenamefont {Zhang}\ \emph {et~al.}(2016)\citenamefont {Zhang},
  \citenamefont {Xia}, \citenamefont {Zhou}, \citenamefont {Wang},
  \citenamefont {Liu}, \citenamefont {Zhao},\ and\ \citenamefont
  {Ezawa}}]{Zhang2016}%
  \BibitemOpen
  \bibfield  {author} {\bibinfo {author} {\bibfnamefont {X.}~\bibnamefont
  {Zhang}}, \bibinfo {author} {\bibfnamefont {J.}~\bibnamefont {Xia}}, \bibinfo
  {author} {\bibfnamefont {Y.}~\bibnamefont {Zhou}}, \bibinfo {author}
  {\bibfnamefont {D.}~\bibnamefont {Wang}}, \bibinfo {author} {\bibfnamefont
  {X.}~\bibnamefont {Liu}}, \bibinfo {author} {\bibfnamefont {W.}~\bibnamefont
  {Zhao}}, \ and\ \bibinfo {author} {\bibfnamefont {M.}~\bibnamefont {Ezawa}},\
  }\href@noop {} {\bibfield  {journal} {\bibinfo  {journal} {Physical Review
  B}\ }\textbf {\bibinfo {volume} {94}},\ \bibinfo {pages} {094420} (\bibinfo
  {year} {2016})}\BibitemShut {NoStop}%
\bibitem [{\citenamefont {Nagaosa}\ and\ \citenamefont
  {Tokura}(2013)}]{Nagaosa2013}%
  \BibitemOpen
  \bibfield  {author} {\bibinfo {author} {\bibfnamefont {N.}~\bibnamefont
  {Nagaosa}}\ and\ \bibinfo {author} {\bibfnamefont {Y.}~\bibnamefont
  {Tokura}},\ }\href {\doibase 10.1038/nnano.2013.243} {\bibfield  {journal}
  {\bibinfo  {journal} {Nature nanotechnology}\ }\textbf {\bibinfo {volume}
  {8}},\ \bibinfo {pages} {899} (\bibinfo {year} {2013})}\BibitemShut {NoStop}%
\bibitem [{\citenamefont {Finazzi}\ \emph {et~al.}(2013)\citenamefont
  {Finazzi}, \citenamefont {Savoini}, \citenamefont {Khorsand}, \citenamefont
  {Tsukamoto}, \citenamefont {Itoh}, \citenamefont {Du{\`{o}}}, \citenamefont
  {Kirilyuk}, \citenamefont {Rasing},\ and\ \citenamefont
  {Ezawa}}]{Finazzi2013}%
  \BibitemOpen
  \bibfield  {author} {\bibinfo {author} {\bibfnamefont {M.}~\bibnamefont
  {Finazzi}}, \bibinfo {author} {\bibfnamefont {M.}~\bibnamefont {Savoini}},
  \bibinfo {author} {\bibfnamefont {A.~R.}\ \bibnamefont {Khorsand}}, \bibinfo
  {author} {\bibfnamefont {A.}~\bibnamefont {Tsukamoto}}, \bibinfo {author}
  {\bibfnamefont {A.}~\bibnamefont {Itoh}}, \bibinfo {author} {\bibfnamefont
  {L.}~\bibnamefont {Du{\`{o}}}}, \bibinfo {author} {\bibfnamefont
  {A.}~\bibnamefont {Kirilyuk}}, \bibinfo {author} {\bibfnamefont
  {T.}~\bibnamefont {Rasing}}, \ and\ \bibinfo {author} {\bibfnamefont
  {M.}~\bibnamefont {Ezawa}},\ }\href {\doibase 10.1103/PhysRevLett.110.177205}
  {\bibfield  {journal} {\bibinfo  {journal} {Physical Review Letters}\
  }\textbf {\bibinfo {volume} {110}},\ \bibinfo {pages} {177205} (\bibinfo
  {year} {2013})}\BibitemShut {NoStop}%
\bibitem [{\citenamefont {Zhang}\ \emph {et~al.}(2018)\citenamefont {Zhang},
  \citenamefont {Kronast}, \citenamefont {van~der Laan},\ and\ \citenamefont
  {Hesjedal}}]{Zhang2018}%
  \BibitemOpen
  \bibfield  {author} {\bibinfo {author} {\bibfnamefont {S.}~\bibnamefont
  {Zhang}}, \bibinfo {author} {\bibfnamefont {F.}~\bibnamefont {Kronast}},
  \bibinfo {author} {\bibfnamefont {G.}~\bibnamefont {van~der Laan}}, \ and\
  \bibinfo {author} {\bibfnamefont {T.}~\bibnamefont {Hesjedal}},\ }\href
  {\doibase 10.1021/acs.nanolett.7b04537} {\bibfield  {journal} {\bibinfo
  {journal} {Nano Letters}\ }\textbf {\bibinfo {volume} {18}},\ \bibinfo
  {pages} {1057} (\bibinfo {year} {2018})}\BibitemShut {NoStop}%
\bibitem [{\citenamefont {Zheng}\ \emph {et~al.}(2017)\citenamefont {Zheng},
  \citenamefont {Li}, \citenamefont {Wang}, \citenamefont {Song}, \citenamefont
  {Jin}, \citenamefont {Wei}, \citenamefont {Kov\'acs}, \citenamefont {Zang},
  \citenamefont {Tian}, \citenamefont {Zhang}, \citenamefont {Du},\ and\
  \citenamefont {Dunin-Borkowski}}]{Zheng2017}%
  \BibitemOpen
  \bibfield  {author} {\bibinfo {author} {\bibfnamefont {F.}~\bibnamefont
  {Zheng}}, \bibinfo {author} {\bibfnamefont {H.}~\bibnamefont {Li}}, \bibinfo
  {author} {\bibfnamefont {S.}~\bibnamefont {Wang}}, \bibinfo {author}
  {\bibfnamefont {D.}~\bibnamefont {Song}}, \bibinfo {author} {\bibfnamefont
  {C.}~\bibnamefont {Jin}}, \bibinfo {author} {\bibfnamefont {W.}~\bibnamefont
  {Wei}}, \bibinfo {author} {\bibfnamefont {A.}~\bibnamefont {Kov\'acs}},
  \bibinfo {author} {\bibfnamefont {J.}~\bibnamefont {Zang}}, \bibinfo {author}
  {\bibfnamefont {M.}~\bibnamefont {Tian}}, \bibinfo {author} {\bibfnamefont
  {Y.}~\bibnamefont {Zhang}}, \bibinfo {author} {\bibfnamefont
  {H.}~\bibnamefont {Du}}, \ and\ \bibinfo {author} {\bibfnamefont {R.~E.}\
  \bibnamefont {Dunin-Borkowski}},\ }\href {\doibase
  10.1103/PhysRevLett.119.197205} {\bibfield  {journal} {\bibinfo  {journal}
  {Physical Review Letters}\ }\textbf {\bibinfo {volume} {119}},\ \bibinfo
  {pages} {197205} (\bibinfo {year} {2017})}\BibitemShut {NoStop}%
\bibitem [{\citenamefont {Jin}\ \emph {et~al.}(2017)\citenamefont {Jin},
  \citenamefont {Li}, \citenamefont {Kov{\'a}cs}, \citenamefont {Caron},
  \citenamefont {Zheng}, \citenamefont {Rybakov}, \citenamefont {Kiselev},
  \citenamefont {Du}, \citenamefont {Bl{\"u}gel}, \citenamefont {Tian} \emph
  {et~al.}}]{Jin2017}%
  \BibitemOpen
  \bibfield  {author} {\bibinfo {author} {\bibfnamefont {C.}~\bibnamefont
  {Jin}}, \bibinfo {author} {\bibfnamefont {Z.-A.}\ \bibnamefont {Li}},
  \bibinfo {author} {\bibfnamefont {A.}~\bibnamefont {Kov{\'a}cs}}, \bibinfo
  {author} {\bibfnamefont {J.}~\bibnamefont {Caron}}, \bibinfo {author}
  {\bibfnamefont {F.}~\bibnamefont {Zheng}}, \bibinfo {author} {\bibfnamefont
  {F.~N.}\ \bibnamefont {Rybakov}}, \bibinfo {author} {\bibfnamefont {N.~S.}\
  \bibnamefont {Kiselev}}, \bibinfo {author} {\bibfnamefont {H.}~\bibnamefont
  {Du}}, \bibinfo {author} {\bibfnamefont {S.}~\bibnamefont {Bl{\"u}gel}},
  \bibinfo {author} {\bibfnamefont {M.}~\bibnamefont {Tian}},  \emph {et~al.},\
  }\href {\doibase doi:10.1038/ncomms15569} {\bibfield  {journal} {\bibinfo
  {journal} {Nature Communications}\ }\textbf {\bibinfo {volume} {8}} (\bibinfo
  {year} {2017}),\ doi:10.1038/ncomms15569}\BibitemShut {NoStop}%
\bibitem [{\citenamefont {Ho}\ \emph {et~al.}(2019)\citenamefont {Ho},
  \citenamefont {Tan}, \citenamefont {Goolaup}, \citenamefont {Oyarce},
  \citenamefont {Raju}, \citenamefont {Huang}, \citenamefont
  {Soumyanarayanan},\ and\ \citenamefont {Panagopoulos}}]{Ho2019}%
  \BibitemOpen
  \bibfield  {author} {\bibinfo {author} {\bibfnamefont {P.}~\bibnamefont
  {Ho}}, \bibinfo {author} {\bibfnamefont {A.~K.}\ \bibnamefont {Tan}},
  \bibinfo {author} {\bibfnamefont {S.}~\bibnamefont {Goolaup}}, \bibinfo
  {author} {\bibfnamefont {A.~G.}\ \bibnamefont {Oyarce}}, \bibinfo {author}
  {\bibfnamefont {M.}~\bibnamefont {Raju}}, \bibinfo {author} {\bibfnamefont
  {L.}~\bibnamefont {Huang}}, \bibinfo {author} {\bibfnamefont
  {A.}~\bibnamefont {Soumyanarayanan}}, \ and\ \bibinfo {author} {\bibfnamefont
  {C.}~\bibnamefont {Panagopoulos}},\ }\href {\doibase
  10.1103/PhysRevApplied.11.024064} {\bibfield  {journal} {\bibinfo  {journal}
  {Phys. Rev. Applied}\ }\textbf {\bibinfo {volume} {11}},\ \bibinfo {pages}
  {024064} (\bibinfo {year} {2019})}\BibitemShut {NoStop}%
\bibitem [{\citenamefont {Fert}\ \emph {et~al.}(2017)\citenamefont {Fert},
  \citenamefont {Reyren},\ and\ \citenamefont {Cros}}]{Fert2017}%
  \BibitemOpen
  \bibfield  {author} {\bibinfo {author} {\bibfnamefont {A.}~\bibnamefont
  {Fert}}, \bibinfo {author} {\bibfnamefont {N.}~\bibnamefont {Reyren}}, \ and\
  \bibinfo {author} {\bibfnamefont {V.}~\bibnamefont {Cros}},\ }\href@noop {}
  {\bibfield  {journal} {\bibinfo  {journal} {Nature Reviews Materials}\
  }\textbf {\bibinfo {volume} {2}},\ \bibinfo {pages} {1} (\bibinfo {year}
  {2017})}\BibitemShut {NoStop}%
\bibitem [{\citenamefont {Bessarab}\ \emph {et~al.}(2015)\citenamefont
  {Bessarab}, \citenamefont {Uzdin},\ and\ \citenamefont
  {J{\'{o}}nsson}}]{Bessarab2015}%
  \BibitemOpen
  \bibfield  {author} {\bibinfo {author} {\bibfnamefont {P.~F.}\ \bibnamefont
  {Bessarab}}, \bibinfo {author} {\bibfnamefont {V.~M.}\ \bibnamefont {Uzdin}},
  \ and\ \bibinfo {author} {\bibfnamefont {H.}~\bibnamefont {J{\'{o}}nsson}},\
  }\href {\doibase 10.1016/j.cpc.2015.07.001} {\bibfield  {journal} {\bibinfo
  {journal} {Computer Physics Communications}\ }\textbf {\bibinfo {volume}
  {196}},\ \bibinfo {pages} {1} (\bibinfo {year} {2015})}\BibitemShut {NoStop}%
\bibitem [{\citenamefont {Cort{\'e}s-Ortu{\~n}o}\ \emph
  {et~al.}(2017)\citenamefont {Cort{\'e}s-Ortu{\~n}o}, \citenamefont {Wang},
  \citenamefont {Beg}, \citenamefont {Pepper}, \citenamefont {Bisotti},
  \citenamefont {Carey}, \citenamefont {Vousden}, \citenamefont {Kluyver},
  \citenamefont {Hovorka},\ and\ \citenamefont {Fangohr}}]{Cortes2017}%
  \BibitemOpen
  \bibfield  {author} {\bibinfo {author} {\bibfnamefont {D.}~\bibnamefont
  {Cort{\'e}s-Ortu{\~n}o}}, \bibinfo {author} {\bibfnamefont {W.}~\bibnamefont
  {Wang}}, \bibinfo {author} {\bibfnamefont {M.}~\bibnamefont {Beg}}, \bibinfo
  {author} {\bibfnamefont {R.~A.}\ \bibnamefont {Pepper}}, \bibinfo {author}
  {\bibfnamefont {M.-A.}\ \bibnamefont {Bisotti}}, \bibinfo {author}
  {\bibfnamefont {R.}~\bibnamefont {Carey}}, \bibinfo {author} {\bibfnamefont
  {M.}~\bibnamefont {Vousden}}, \bibinfo {author} {\bibfnamefont
  {T.}~\bibnamefont {Kluyver}}, \bibinfo {author} {\bibfnamefont
  {O.}~\bibnamefont {Hovorka}}, \ and\ \bibinfo {author} {\bibfnamefont
  {H.}~\bibnamefont {Fangohr}},\ }\href {\doibase 10.1038/s41598-017-03391-8}
  {\bibfield  {journal} {\bibinfo  {journal} {Scientific Reports}\ }\textbf
  {\bibinfo {volume} {7}},\ \bibinfo {pages} {4060} (\bibinfo {year}
  {2017})}\BibitemShut {NoStop}%
\bibitem [{\citenamefont {Stosic}\ \emph {et~al.}(2017)\citenamefont {Stosic},
  \citenamefont {Mulkers}, \citenamefont {Van~Waeyenberge}, \citenamefont
  {Ludermir},\ and\ \citenamefont {Milo\ifmmode \check{s}\else
  \v{s}\fi{}evi\ifmmode~\acute{c}\else \'{c}\fi{}}}]{Stosic2017}%
  \BibitemOpen
  \bibfield  {author} {\bibinfo {author} {\bibfnamefont {D.}~\bibnamefont
  {Stosic}}, \bibinfo {author} {\bibfnamefont {J.}~\bibnamefont {Mulkers}},
  \bibinfo {author} {\bibfnamefont {B.}~\bibnamefont {Van~Waeyenberge}},
  \bibinfo {author} {\bibfnamefont {T.~B.}\ \bibnamefont {Ludermir}}, \ and\
  \bibinfo {author} {\bibfnamefont {M.~V.}\ \bibnamefont {Milo\ifmmode
  \check{s}\else \v{s}\fi{}evi\ifmmode~\acute{c}\else \'{c}\fi{}}},\ }\href
  {\doibase 10.1103/PhysRevB.95.214418} {\bibfield  {journal} {\bibinfo
  {journal} {Physical Review B}\ }\textbf {\bibinfo {volume} {95}},\ \bibinfo
  {pages} {214418} (\bibinfo {year} {2017})}\BibitemShut {NoStop}%
\bibitem [{\citenamefont {Bessarab}\ \emph {et~al.}(2018)\citenamefont
  {Bessarab}, \citenamefont {M{\"u}ller}, \citenamefont {Lobanov},
  \citenamefont {Rybakov}, \citenamefont {Kiselev}, \citenamefont
  {J{\'o}nsson}, \citenamefont {Uzdin}, \citenamefont {Bl{\"u}gel},
  \citenamefont {Bergqvist},\ and\ \citenamefont {Delin}}]{Bessarab2018}%
  \BibitemOpen
  \bibfield  {author} {\bibinfo {author} {\bibfnamefont {P.~F.}\ \bibnamefont
  {Bessarab}}, \bibinfo {author} {\bibfnamefont {G.~P.}\ \bibnamefont
  {M{\"u}ller}}, \bibinfo {author} {\bibfnamefont {I.~S.}\ \bibnamefont
  {Lobanov}}, \bibinfo {author} {\bibfnamefont {F.~N.}\ \bibnamefont
  {Rybakov}}, \bibinfo {author} {\bibfnamefont {N.~S.}\ \bibnamefont
  {Kiselev}}, \bibinfo {author} {\bibfnamefont {H.}~\bibnamefont
  {J{\'o}nsson}}, \bibinfo {author} {\bibfnamefont {V.~M.}\ \bibnamefont
  {Uzdin}}, \bibinfo {author} {\bibfnamefont {S.}~\bibnamefont {Bl{\"u}gel}},
  \bibinfo {author} {\bibfnamefont {L.}~\bibnamefont {Bergqvist}}, \ and\
  \bibinfo {author} {\bibfnamefont {A.}~\bibnamefont {Delin}},\ }\href
  {\doibase 10.1038/s41598-018-21623-3} {\bibfield  {journal} {\bibinfo
  {journal} {Scientific Reports}\ }\textbf {\bibinfo {volume} {8}},\ \bibinfo
  {pages} {3433} (\bibinfo {year} {2018})}\BibitemShut {NoStop}%
\bibitem [{\citenamefont {Wiesendanger}(2009)}]{Wiesendanger2009}%
  \BibitemOpen
  \bibfield  {author} {\bibinfo {author} {\bibfnamefont {R.}~\bibnamefont
  {Wiesendanger}},\ }\href {\doibase 10.1103/RevModPhys.81.1495} {\bibfield
  {journal} {\bibinfo  {journal} {Rev. Mod. Phys.}\ }\textbf {\bibinfo {volume}
  {81}},\ \bibinfo {pages} {1495} (\bibinfo {year} {2009})}\BibitemShut
  {NoStop}%
\bibitem [{sup()}]{suppmat}%
  \BibitemOpen
  \href@noop {} {}\bibinfo {note} {{See
  \href{https://arxiv.org/src/1901.06999/anc/supplementary_material.pdf}{\color{blue}{Supplemental
  Material.}}}}\BibitemShut {Stop}%
\bibitem [{\citenamefont {Hanneken}\ \emph {et~al.}(2015)\citenamefont
  {Hanneken}, \citenamefont {Otte}, \citenamefont {Kubetzka}, \citenamefont
  {Dup{\'e}}, \citenamefont {Romming}, \citenamefont {von Bergmann},
  \citenamefont {Wiesendanger},\ and\ \citenamefont {Heinze}}]{Hanneken2015}%
  \BibitemOpen
  \bibfield  {author} {\bibinfo {author} {\bibfnamefont {C.}~\bibnamefont
  {Hanneken}}, \bibinfo {author} {\bibfnamefont {F.}~\bibnamefont {Otte}},
  \bibinfo {author} {\bibfnamefont {A.}~\bibnamefont {Kubetzka}}, \bibinfo
  {author} {\bibfnamefont {B.}~\bibnamefont {Dup{\'e}}}, \bibinfo {author}
  {\bibfnamefont {N.}~\bibnamefont {Romming}}, \bibinfo {author} {\bibfnamefont
  {K.}~\bibnamefont {von Bergmann}}, \bibinfo {author} {\bibfnamefont
  {R.}~\bibnamefont {Wiesendanger}}, \ and\ \bibinfo {author} {\bibfnamefont
  {S.}~\bibnamefont {Heinze}},\ }\href
  {http://dx.doi.org/10.1038/nnano.2015.218} {\bibfield  {journal} {\bibinfo
  {journal} {Nature Nanotechnology}\ }\textbf {\bibinfo {volume} {10}},\
  \bibinfo {pages} {1039} (\bibinfo {year} {2015})}\BibitemShut {NoStop}%
\bibitem [{\citenamefont {Bisotti}\ \emph {et~al.}(2018)\citenamefont
  {Bisotti}, \citenamefont {Cort{\'e}s-Ortu{\~n}o}, \citenamefont {Pepper},
  \citenamefont {Wang}, \citenamefont {Beg}, \citenamefont {Kluyver},\ and\
  \citenamefont {Fangohr}}]{Bisotti2018}%
  \BibitemOpen
  \bibfield  {author} {\bibinfo {author} {\bibfnamefont {M.-A.}\ \bibnamefont
  {Bisotti}}, \bibinfo {author} {\bibfnamefont {D.}~\bibnamefont
  {Cort{\'e}s-Ortu{\~n}o}}, \bibinfo {author} {\bibfnamefont {R.}~\bibnamefont
  {Pepper}}, \bibinfo {author} {\bibfnamefont {W.}~\bibnamefont {Wang}},
  \bibinfo {author} {\bibfnamefont {M.}~\bibnamefont {Beg}}, \bibinfo {author}
  {\bibfnamefont {T.}~\bibnamefont {Kluyver}}, \ and\ \bibinfo {author}
  {\bibfnamefont {H.}~\bibnamefont {Fangohr}},\ }\href {\doibase
  10.5334/jors.223} {\bibfield  {journal} {\bibinfo  {journal} {Journal of Open
  Research Software}\ }\textbf {\bibinfo {volume} {6}},\ \bibinfo {pages} {22}
  (\bibinfo {year} {2018})}\BibitemShut {NoStop}%
\bibitem [{\citenamefont {Leonov}\ \emph {et~al.}(2016)\citenamefont {Leonov},
  \citenamefont {Monchesky}, \citenamefont {Romming}, \citenamefont {Kubetzka},
  \citenamefont {Bogdanov},\ and\ \citenamefont {Wiesendanger}}]{Leonov2016}%
  \BibitemOpen
  \bibfield  {author} {\bibinfo {author} {\bibfnamefont {A.~O.}\ \bibnamefont
  {Leonov}}, \bibinfo {author} {\bibfnamefont {T.~L.}\ \bibnamefont
  {Monchesky}}, \bibinfo {author} {\bibfnamefont {N.}~\bibnamefont {Romming}},
  \bibinfo {author} {\bibfnamefont {A.}~\bibnamefont {Kubetzka}}, \bibinfo
  {author} {\bibfnamefont {A.~N.}\ \bibnamefont {Bogdanov}}, \ and\ \bibinfo
  {author} {\bibfnamefont {R.}~\bibnamefont {Wiesendanger}},\ }\href {\doibase
  10.1088/1367-2630/18/6/065003} {\bibfield  {journal} {\bibinfo  {journal}
  {New Journal of Physics}\ }\textbf {\bibinfo {volume} {18}},\ \bibinfo
  {pages} {065003} (\bibinfo {year} {2016})}\BibitemShut {NoStop}%
\bibitem [{\citenamefont {Lobanov}\ \emph {et~al.}(2016)\citenamefont
  {Lobanov}, \citenamefont {J\'onsson},\ and\ \citenamefont
  {Uzdin}}]{Lobanov2016}%
  \BibitemOpen
  \bibfield  {author} {\bibinfo {author} {\bibfnamefont {I.~S.}\ \bibnamefont
  {Lobanov}}, \bibinfo {author} {\bibfnamefont {H.}~\bibnamefont {J\'onsson}},
  \ and\ \bibinfo {author} {\bibfnamefont {V.~M.}\ \bibnamefont {Uzdin}},\
  }\href {\doibase 10.1103/PhysRevB.94.174418} {\bibfield  {journal} {\bibinfo
  {journal} {Physical Review B}\ }\textbf {\bibinfo {volume} {94}},\ \bibinfo
  {pages} {174418} (\bibinfo {year} {2016})}\BibitemShut {NoStop}%
\bibitem [{\citenamefont {Wang}\ \emph {et~al.}(2018)\citenamefont {Wang},
  \citenamefont {Yuan},\ and\ \citenamefont {Wang}}]{Wang2018}%
  \BibitemOpen
  \bibfield  {author} {\bibinfo {author} {\bibfnamefont {X.~S.}\ \bibnamefont
  {Wang}}, \bibinfo {author} {\bibfnamefont {H.~Y.}\ \bibnamefont {Yuan}}, \
  and\ \bibinfo {author} {\bibfnamefont {X.~R.}\ \bibnamefont {Wang}},\ }\href
  {\doibase 10.1038/s42005-018-0029-0} {\bibfield  {journal} {\bibinfo
  {journal} {Communications Physics}\ }\textbf {\bibinfo {volume} {1}},\
  \bibinfo {pages} {31} (\bibinfo {year} {2018})}\BibitemShut {NoStop}%
\bibitem [{\citenamefont {von Malottki}\ \emph {et~al.}(2017)\citenamefont {von
  Malottki}, \citenamefont {Dup{\'e}}, \citenamefont {Bessarab}, \citenamefont
  {Delin},\ and\ \citenamefont {Heinze}}]{Malottki2017}%
  \BibitemOpen
  \bibfield  {author} {\bibinfo {author} {\bibfnamefont {S.}~\bibnamefont {von
  Malottki}}, \bibinfo {author} {\bibfnamefont {B.}~\bibnamefont {Dup{\'e}}},
  \bibinfo {author} {\bibfnamefont {P.~F.}\ \bibnamefont {Bessarab}}, \bibinfo
  {author} {\bibfnamefont {A.}~\bibnamefont {Delin}}, \ and\ \bibinfo {author}
  {\bibfnamefont {S.}~\bibnamefont {Heinze}},\ }\href {\doibase
  10.1038/s41598-017-12525-x} {\bibfield  {journal} {\bibinfo  {journal}
  {Scientific Reports}\ }\textbf {\bibinfo {volume} {7}},\ \bibinfo {pages}
  {12299} (\bibinfo {year} {2017})}\BibitemShut {NoStop}%
\bibitem [{\citenamefont {Kiselev}\ \emph {et~al.}(2011)\citenamefont
  {Kiselev}, \citenamefont {Bogdanov}, \citenamefont {Sch\"{a}fer},\ and\
  \citenamefont {R\"{o}{\ss}ler}}]{Kiselev2011}%
  \BibitemOpen
  \bibfield  {author} {\bibinfo {author} {\bibfnamefont {N.~S.}\ \bibnamefont
  {Kiselev}}, \bibinfo {author} {\bibfnamefont {A.~N.}\ \bibnamefont
  {Bogdanov}}, \bibinfo {author} {\bibfnamefont {R.}~\bibnamefont
  {Sch\"{a}fer}}, \ and\ \bibinfo {author} {\bibfnamefont {U.~K.}\ \bibnamefont
  {R\"{o}{\ss}ler}},\ }\href {\doibase 10.1088/0022-3727/44/39/392001}
  {\bibfield  {journal} {\bibinfo  {journal} {Journal of Physics D: Applied
  Physics}\ }\textbf {\bibinfo {volume} {44}},\ \bibinfo {pages} {392001}
  (\bibinfo {year} {2011})}\BibitemShut {NoStop}%
\bibitem [{\citenamefont {Berg}\ and\ \citenamefont
  {L\"{u}scher}(1981)}]{Berg1981}%
  \BibitemOpen
  \bibfield  {author} {\bibinfo {author} {\bibfnamefont {B.}~\bibnamefont
  {Berg}}\ and\ \bibinfo {author} {\bibfnamefont {M.}~\bibnamefont
  {L\"{u}scher}},\ }\href {\doibase 10.1016/0550-3213(81)90568-X} {\bibfield
  {journal} {\bibinfo  {journal} {Nuclear Physics, Section B}\ }\textbf
  {\bibinfo {volume} {190}},\ \bibinfo {pages} {412} (\bibinfo {year}
  {1981})}\BibitemShut {NoStop}%
\bibitem [{\citenamefont {Yin}\ \emph {et~al.}(2016)\citenamefont {Yin},
  \citenamefont {Li}, \citenamefont {Kong}, \citenamefont {Lake}, \citenamefont
  {Chien},\ and\ \citenamefont {Zang}}]{Yin2016}%
  \BibitemOpen
  \bibfield  {author} {\bibinfo {author} {\bibfnamefont {G.}~\bibnamefont
  {Yin}}, \bibinfo {author} {\bibfnamefont {Y.}~\bibnamefont {Li}}, \bibinfo
  {author} {\bibfnamefont {L.}~\bibnamefont {Kong}}, \bibinfo {author}
  {\bibfnamefont {R.~K.}\ \bibnamefont {Lake}}, \bibinfo {author}
  {\bibfnamefont {C.~L.}\ \bibnamefont {Chien}}, \ and\ \bibinfo {author}
  {\bibfnamefont {J.}~\bibnamefont {Zang}},\ }\href {\doibase
  10.1103/PhysRevB.93.174403} {\bibfield  {journal} {\bibinfo  {journal}
  {Physical Review B}\ }\textbf {\bibinfo {volume} {93}},\ \bibinfo {pages}
  {174403} (\bibinfo {year} {2016})}\BibitemShut {NoStop}%
\bibitem [{\citenamefont {Bogdanov}\ and\ \citenamefont
  {Hubert}(1994{\natexlab{b}})}]{Bogdanov1994a}%
  \BibitemOpen
  \bibfield  {author} {\bibinfo {author} {\bibfnamefont {A.}~\bibnamefont
  {Bogdanov}}\ and\ \bibinfo {author} {\bibfnamefont {A.}~\bibnamefont
  {Hubert}},\ }\href {\doibase 10.1002/pssb.2221860223} {\bibfield  {journal}
  {\bibinfo  {journal} {physica status solidi (b)}\ }\textbf {\bibinfo {volume}
  {186}},\ \bibinfo {pages} {527} (\bibinfo {year}
  {1994}{\natexlab{b}})}\BibitemShut {NoStop}%
\bibitem [{\citenamefont {Hsu}\ \emph {et~al.}(2016)\citenamefont {Hsu},
  \citenamefont {Kubetzka}, \citenamefont {Finco}, \citenamefont {Romming},
  \citenamefont {von Bergmann},\ and\ \citenamefont {Wiesendanger}}]{Hsu2016}%
  \BibitemOpen
  \bibfield  {author} {\bibinfo {author} {\bibfnamefont {P.-J.}\ \bibnamefont
  {Hsu}}, \bibinfo {author} {\bibfnamefont {A.}~\bibnamefont {Kubetzka}},
  \bibinfo {author} {\bibfnamefont {A.}~\bibnamefont {Finco}}, \bibinfo
  {author} {\bibfnamefont {N.}~\bibnamefont {Romming}}, \bibinfo {author}
  {\bibfnamefont {K.}~\bibnamefont {von Bergmann}}, \ and\ \bibinfo {author}
  {\bibfnamefont {R.}~\bibnamefont {Wiesendanger}},\ }\href
  {https://doi.org/10.1038/nnano.2016.234} {\bibfield  {journal} {\bibinfo
  {journal} {Nature Nanotechnology}\ }\textbf {\bibinfo {volume} {12}},\
  \bibinfo {pages} {123} (\bibinfo {year} {2016})}\BibitemShut {NoStop}%
\bibitem [{\citenamefont {Mulkers}\ \emph {et~al.}(2016)\citenamefont
  {Mulkers}, \citenamefont {Milo\ifmmode \check{s}\else
  \v{s}\fi{}evi\ifmmode~\acute{c}\else \'{c}\fi{}},\ and\ \citenamefont
  {Van~Waeyenberge}}]{Mulkers2016}%
  \BibitemOpen
  \bibfield  {author} {\bibinfo {author} {\bibfnamefont {J.}~\bibnamefont
  {Mulkers}}, \bibinfo {author} {\bibfnamefont {M.~V.}\ \bibnamefont
  {Milo\ifmmode \check{s}\else \v{s}\fi{}evi\ifmmode~\acute{c}\else
  \'{c}\fi{}}}, \ and\ \bibinfo {author} {\bibfnamefont {B.}~\bibnamefont
  {Van~Waeyenberge}},\ }\href {\doibase 10.1103/PhysRevB.93.214405} {\bibfield
  {journal} {\bibinfo  {journal} {Phys. Rev. B}\ }\textbf {\bibinfo {volume}
  {93}},\ \bibinfo {pages} {214405} (\bibinfo {year} {2016})}\BibitemShut
  {NoStop}%
\bibitem [{\citenamefont {Zhao}\ \emph {et~al.}(2016)\citenamefont {Zhao},
  \citenamefont {Jin}, \citenamefont {Wang}, \citenamefont {Du}, \citenamefont
  {Zang}, \citenamefont {Tian}, \citenamefont {Che},\ and\ \citenamefont
  {Zhang}}]{Zhao2016}%
  \BibitemOpen
  \bibfield  {author} {\bibinfo {author} {\bibfnamefont {X.}~\bibnamefont
  {Zhao}}, \bibinfo {author} {\bibfnamefont {C.}~\bibnamefont {Jin}}, \bibinfo
  {author} {\bibfnamefont {C.}~\bibnamefont {Wang}}, \bibinfo {author}
  {\bibfnamefont {H.}~\bibnamefont {Du}}, \bibinfo {author} {\bibfnamefont
  {J.}~\bibnamefont {Zang}}, \bibinfo {author} {\bibfnamefont {M.}~\bibnamefont
  {Tian}}, \bibinfo {author} {\bibfnamefont {R.}~\bibnamefont {Che}}, \ and\
  \bibinfo {author} {\bibfnamefont {Y.}~\bibnamefont {Zhang}},\ }\href
  {\doibase 10.1073/pnas.1600197113} {\bibfield  {journal} {\bibinfo  {journal}
  {Proceedings of the National Academy of Sciences}\ }\textbf {\bibinfo
  {volume} {113}},\ \bibinfo {pages} {4918} (\bibinfo {year}
  {2016})}\BibitemShut {NoStop}%
\bibitem [{\citenamefont {Foster}\ \emph {et~al.}(2019)\citenamefont {Foster},
  \citenamefont {Kind}, \citenamefont {Ackerman}, \citenamefont {Tai},
  \citenamefont {Dennis},\ and\ \citenamefont {Smalyukh}}]{Foster2019}%
  \BibitemOpen
  \bibfield  {author} {\bibinfo {author} {\bibfnamefont {D.}~\bibnamefont
  {Foster}}, \bibinfo {author} {\bibfnamefont {C.}~\bibnamefont {Kind}},
  \bibinfo {author} {\bibfnamefont {P.~J.}\ \bibnamefont {Ackerman}}, \bibinfo
  {author} {\bibfnamefont {J.-S.~B.}\ \bibnamefont {Tai}}, \bibinfo {author}
  {\bibfnamefont {M.~R.}\ \bibnamefont {Dennis}}, \ and\ \bibinfo {author}
  {\bibfnamefont {I.~I.}\ \bibnamefont {Smalyukh}},\ }\href {\doibase
  10.1038/s41567-019-0476-x} {\bibfield  {journal} {\bibinfo  {journal} {Nature
  Physics}\ } (\bibinfo {year} {2019}),\ 10.1038/s41567-019-0476-x}\BibitemShut
  {NoStop}%
\bibitem [{\citenamefont {Rybakov}\ and\ \citenamefont
  {Kiselev}(2019)}]{Rybakov2019}%
  \BibitemOpen
  \bibfield  {author} {\bibinfo {author} {\bibfnamefont {F.~N.}\ \bibnamefont
  {Rybakov}}\ and\ \bibinfo {author} {\bibfnamefont {N.~S.}\ \bibnamefont
  {Kiselev}},\ }\href {\doibase 10.1103/PhysRevB.99.064437} {\bibfield
  {journal} {\bibinfo  {journal} {Phys. Rev. B}\ }\textbf {\bibinfo {volume}
  {99}},\ \bibinfo {pages} {064437} (\bibinfo {year} {2019})}\BibitemShut
  {NoStop}%
\bibitem [{\citenamefont {von Malottki}\ \emph {et~al.}(2019)\citenamefont {von
  Malottki}, \citenamefont {Bessarab}, \citenamefont {Haldar}, \citenamefont
  {Delin},\ and\ \citenamefont {Heinze}}]{Malottki2019}%
  \BibitemOpen
  \bibfield  {author} {\bibinfo {author} {\bibfnamefont {S.}~\bibnamefont {von
  Malottki}}, \bibinfo {author} {\bibfnamefont {P.~F.}\ \bibnamefont
  {Bessarab}}, \bibinfo {author} {\bibfnamefont {S.}~\bibnamefont {Haldar}},
  \bibinfo {author} {\bibfnamefont {A.}~\bibnamefont {Delin}}, \ and\ \bibinfo
  {author} {\bibfnamefont {S.}~\bibnamefont {Heinze}},\ }\href {\doibase
  10.1103/PhysRevB.99.060409} {\bibfield  {journal} {\bibinfo  {journal} {Phys.
  Rev. B}\ }\textbf {\bibinfo {volume} {99}},\ \bibinfo {pages} {060409(R)}
  (\bibinfo {year} {2019})}\BibitemShut {NoStop}%
\bibitem [{\citenamefont {Dup{\'{e}}}\ \emph {et~al.}(2014)\citenamefont
  {Dup{\'{e}}}, \citenamefont {Hoffmann}, \citenamefont {Paillard},\ and\
  \citenamefont {Heinze}}]{Dupe2014}%
  \BibitemOpen
  \bibfield  {author} {\bibinfo {author} {\bibfnamefont {B.}~\bibnamefont
  {Dup{\'{e}}}}, \bibinfo {author} {\bibfnamefont {M.}~\bibnamefont
  {Hoffmann}}, \bibinfo {author} {\bibfnamefont {C.}~\bibnamefont {Paillard}},
  \ and\ \bibinfo {author} {\bibfnamefont {S.}~\bibnamefont {Heinze}},\ }\href
  {\doibase 10.1038/ncomms5030} {\bibfield  {journal} {\bibinfo  {journal}
  {Nature Communications}\ }\textbf {\bibinfo {volume} {5}},\ \bibinfo {pages}
  {4030} (\bibinfo {year} {2014})}\BibitemShut {NoStop}%
\bibitem [{\citenamefont {{Bouhassoune}}\ \emph {et~al.}(2019)\citenamefont
  {{Bouhassoune}}, \citenamefont {{Fernandes}}, \citenamefont {{Bl{\"u}gel}},\
  and\ \citenamefont {{Lounis}}}]{Bouhassoune2019}%
  \BibitemOpen
  \bibfield  {author} {\bibinfo {author} {\bibfnamefont {M.}~\bibnamefont
  {{Bouhassoune}}}, \bibinfo {author} {\bibfnamefont {I.~L.}\ \bibnamefont
  {{Fernandes}}}, \bibinfo {author} {\bibfnamefont {S.}~\bibnamefont
  {{Bl{\"u}gel}}}, \ and\ \bibinfo {author} {\bibfnamefont {S.}~\bibnamefont
  {{Lounis}}},\ }\href {https://arxiv.org/abs/1901.07029} {\bibfield  {journal}
  {\bibinfo  {journal} {arXiv e-prints}\ ,\ \bibinfo {pages}
  {arXiv:1901.07029}} (\bibinfo {year} {2019})}\BibitemShut {NoStop}%
\bibitem [{\citenamefont {Cort{\'e}s-Ortu{\~n}o}\ \emph
  {et~al.}(2019)\citenamefont {Cort{\'e}s-Ortu{\~n}o}, \citenamefont {Romming},
  \citenamefont {Beg}, \citenamefont {von Bergmann}, \citenamefont {Kubetzka},
  \citenamefont {Hovorka}, \citenamefont {Fangohr},\ and\ \citenamefont
  {Wiesendanger}}]{Cortes2019}%
  \BibitemOpen
  \bibfield  {author} {\bibinfo {author} {\bibfnamefont {D.}~\bibnamefont
  {Cort{\'e}s-Ortu{\~n}o}}, \bibinfo {author} {\bibfnamefont {N.}~\bibnamefont
  {Romming}}, \bibinfo {author} {\bibfnamefont {M.}~\bibnamefont {Beg}},
  \bibinfo {author} {\bibfnamefont {K.}~\bibnamefont {von Bergmann}}, \bibinfo
  {author} {\bibfnamefont {A.}~\bibnamefont {Kubetzka}}, \bibinfo {author}
  {\bibfnamefont {O.}~\bibnamefont {Hovorka}}, \bibinfo {author} {\bibfnamefont
  {H.}~\bibnamefont {Fangohr}}, \ and\ \bibinfo {author} {\bibfnamefont
  {R.}~\bibnamefont {Wiesendanger}},\ }\href {\doibase 10.5281/zenodo.1438396}
  {\enquote {\bibinfo {title} {{Data set for: Nano-scale magnetic skyrmions and
  target states in confined geometries}},}\ }\bibinfo {howpublished} {Zenodo
  \url{doi:10.5281/zenodo.1438396}. Github:
  \url{https://github.com/davidcortesortuno/paper-2019_nanoscale_skyrmions_target_states_confined_geometries}}
  (\bibinfo {year} {2019})\BibitemShut {NoStop}%
\end{thebibliography}%

\end{document}